\documentclass[12pt]{iopart}

\usepackage{amssymb,braket}
\usepackage[bb=boondox]{mathalfa}
\usepackage{hyperref}
\usepackage{tikz}
\usetikzlibrary{decorations.pathmorphing}
\usetikzlibrary{patterns}
\usetikzlibrary{decorations.markings}
\usetikzlibrary{arrows}
\usetikzlibrary{decorations.pathreplacing}
\usetikzlibrary{calc,intersections,through,backgrounds}
\tikzset{snake it/.style={decorate, decoration={snake,amplitude=.6mm,segment length=1.5mm}}}
\definecolor{dgreen}{rgb}{0,0.4,0}

\newcommand{\ie}{{\it i.e.},\ }

\newcommand{\id}{\mathbb{1}}
\newcommand{\ii}{\mathrm{i}}
\renewcommand{\Tr}{\mathrm{Tr}}

\newcommand{\rr}{\mathbf{r}}

\usepackage{appendix}
\def\duzomniejsze{<\kern-.7mm<}
\def\duzowieksze{>\kern-.7mm>}

\def\textbf#1{{\bf #1}}
\def\beq{\begin{equation}}
\def\eeq{\end{equation}}
\def\be{\begin{equation}}
\def\ee{\end{equation}}
\def\ben{\begin{eqnarray}}
\def\een{\end{eqnarray}}
\def\beqa{\begin{eqnarray}}
\def\eeqa{\end{eqnarray}}
\def\eea{\end{array}}
\def\bea{\begin{array}}
\newcommand{\bei}{\begin{itemize}}
\newcommand{\eei}{\end{itemize}}
\newcommand{\bee}{\begin{enumerate}}
\newcommand{\eee}{\end{enumerate}}

\usepackage{xcolor}
\newcounter{mnotecount}[section]
\renewcommand{\themnotecount}{\thesection.\arabic{mnotecount}}
\newcommand{\mnote}[1]%{}
{\protect{\stepcounter{mnotecount}}$^{\mbox{\footnotesize
$%\!\!\!\!\!\!\,
\bullet$\themnotecount}}$ \marginpar{%\color{red}%
\raggedright\tiny\em
$\!\!\!\!\!\!\,\bullet$\themnotecount: #1} }

\usepackage{soul}

\begin{document}

\title[Average entanglement entropy in a pure Gaussian state ensemble]{Average entanglement entropy of a small subsystem in a constrained pure Gaussian state ensemble}

\author{Erik Aurell}
\address{KTH – Royal Institute of Technology, Alba Nova University Center, SE-106 91 Stockholm, Sweden}

\author{Lucas Hackl}
\address{School of Mathematics and Statistics, The University of Melbourne, Parkville, VIC 3010, Australia}
\address{School of Physics, The University of Melbourne, Parkville, VIC 3010, Australia}

\author{Mario Kieburg}
\address{School of Mathematics and Statistics, The University of Melbourne, Parkville, VIC 3010, Australia}

%\ead{customerservices@ioppublishing.org}
% \vspace{10pt}
% \begin{indented}
% \item[]August 2017 (minor update March 2024)
% \end{indented}

\begin{abstract}
We consider ensembles of pure Gaussian states parametrized by single-mode marginals and (optionally) specific mode-mode correlations. Such ensembles provide a model for the final states when isolated quantum systems thermalize, as they can reproduce thermal properties locally, while being globally pure. By an analysis using real replicas and the coherent state representation of Gaussian states we show that the average entanglement entropy of a small subsystem is the same as the von Neumann entropy of a mixed Gaussian state with the same marginals, but no correlations. Finally, we discuss how these ensembles provide a model for Hawking radiation assuming unitary evolution, and discuss some of their properties in relations to the Page curve of Hawking radiation.
\end{abstract}

%
% Uncomment for keywords
%\vspace{2pc}
%\noindent{\it Keywords}: XXXXXX, YYYYYYYY, ZZZZZZZZZ
%
% Uncomment for Submitted to journal title message
%\submitto{\JPA}
%
% Uncomment if a separate title page is required
%\maketitle
% 
% For two-column output uncomment the next line and choose [10pt] rather than [12pt] in the \documentclass declaration
%\ioptwocol
%

\section{Introduction}
The analysis presented in this manuscript is motivated by two striking observations. The first is Hawking's discovery that black holes emit thermal radiation~\cite{Hawking74}. The second is the eigenstate thermalization hypothesis, as formulated by Deutsch and Srednicki~\cite{deutsch1991quantum,srednicki1994chaos}, which aims to explain why many isolated quantum systems (pure states) are well-described by thermal states (mixed states). In both scenarios, we consider a system that looks thermal from the perspective of individual subsystems, but may be globally in a pure state.

The challenge to understand black holes as quantum objects has held center stage in fundamental physics since the discovery of Hawking radiation, now 50 years ago~\cite{Hawking74,Hawking75,Wald75,wald1994quantum}. Hawking radiation is mode-by-mode thermal (up to "grey body factors"), and in quantum mechanics entropy of an isolated object is conserved.  
Therefore, if the black hole was created from an object of low entropy and the Hawking radiation is all that is left, once the black hole is fully evaporated, either quantum mechanics does not hold in a black hole, or the modes of the Hawking radiation must be entangled, as first observed by Page~\cite{Page1980}. 
This scenario raises the question of how much entangled and entangled in which way.

The thermalization of isolated quantum systems~\cite{deutsch1991quantum,srednicki1994chaos,rigol2008thermalization,polkovnikov2011colloquium,eisert2015quantum} describes the phenomena that unitary evolution of a relatively generic pure initial state yields a late time state, whose properties, such as expectation values of physical observables, are well characterized by a thermal state of a temperature that could have been predicted from the initial energy expectation value. This is a central topic in nonequilibrium quantum statistical mechanics with important implications for quantum information, condensed matter physics, and many-body theory. There has been a tremendous amount of work on identifying different mechanisms, such as the eigenstate thermalization hypothesis, that cause this phenomena and criteria to identify it. Here, we focus on constructing the ensemble of (pure) quantum states compatible with this phenoma using random matrix theory methods.

In a previous contribution \cite{5authors}, we constructed the ensemble of all pure bosonic Gaussian quantum states that share the same marginals. This \emph{compact} set is fully characterized by a set of symplectic transformations subject to marginal constraints and equipped with a natural measure induced by the (non-compact) symplectic group. Put simply, we provided an analytical construction for the set of all states that are globally pure, but look locally thermal in a well-prescribed manner. Applied to Hawking radiation, we showed that in this ensemble of random pure Gaussian states which perfectly mimic the one-mode properties of Hawking radiation, any two modes are almost surely not entangled. The essential large parameter is the total number of excited modes in Hawking radiation, about ${{N}}_\odot \sim 10^{76}$ for a solar-mass black hole.

Here we go further and show that the reduced state of any small subsystem within the aforementioned random pure state ensemble behaves just in the same as mixed state without any correlations between the subsystems. In particular, this implies that such small systems are maximally entangled with the remaining set of modes. When applied to Hawking radiation and consistent with our previous finding, this implies that typical quantum correlations between different small sets of Hawking modes are vanishingly small, but correlations between a small selection of modes and the collective of all other modes is maximal. We make this precise by showing explicitly how the bi-partite entanglement entropy is maximal for small subsystems.

% Here we go further and show that the reduced state of any small subsystem, \ie small number of modes compared to ${{N}}_{M(0)}$, is very nearly thermal in the same random pure state model. $M(0)$ is here the initial mass of the black hole.
% Furthermore, also going beyond \cite{5authors}, we here incorporate Wald's constraint that two modes of Hawking radiation in the same time window are uncorrelated~\cite{Wald75,wald1994quantum}.

\section{Gaussian pure states with fixed marginals}\label{sec:Gaussian-fixed-marginals}
We discuss the ensemble of pure Gaussian states of $N$ modes where we fix all $N$ single mode marginals, \ie each reduction to a single bosonic mode.

\subsection{Marginals and subsystems}
We consider a quantum system with Hilbert space
\begin{equation}
    \mathcal{H}=\bigotimes^N_{i=1}\mathcal{H}_i\,,\label{eq:Hspace}
\end{equation}
where $\mathcal{H}_i$ represent the Hilbert spaces different subsystems. Given a Hamiltonian $\hat{H}=\sum^N_{i=1}\hat{H}_i$, such that $\hat{H}_i$ only acts non-trivially on $\mathcal{H}_i$, we have the thermal state
\begin{equation}
    \rho(\beta)=\frac{1}{Z}e^{-\beta\hat{H}}\quad\mathrm{with}\quad Z=\Tr(e^{-\beta\hat{H}})
\end{equation}
at inverse temperature $\beta=1/T$. When we perform a partial trace over the complement of $\mathcal{H}_i$, that is the $\mathcal{H}_i^c=\otimes_{j\neq i}\mathcal{H}_j$ consisting of all the Hilbert spaces $\mathcal{H}_j$ with $j\neq i$, we recover
\begin{equation}
    \rho_i(\beta)=\Tr_{\mathcal{H}_i^c}\,\rho(\beta)=\frac{1}{Z_i}e^{-\beta\hat{H}_i}\quad\mathrm{with}\quad Z_i=\Tr(e^{-\beta\hat{H}_i})\,.
\end{equation}
These are the marginals of $\rho(\beta)$ and for our simple Hamiltonian $H$ without any coupling between different modes, we find that the individual marginals of $\rho(\beta)$ are themselves thermal states with respect to the subsystem Hamiltonian $\hat{H}_i$.

The general question of how much can be learned about a state from knowing its marginals is known as the quantum marginal problem. In particular, this includes the question if for a given set of (not necessarily disjoint) marginals there exists a state in the full system (mixed or pure) with these marginals. Moreover, if there exists such a state there usually exists a large family, so it becomes also a relevant question on how to parametrize this family and if there is a natural measure over it.

The case of the Gaussian marginal problem for bosons was solved in~\cite{Eisert2008}, which provided sets of inequalities involving the symplectic eigenvalues of the covariance matrix in the whole system and in each subsystem. Various cases and variants have been considered and understood in the literature~\cite{yu2021complete}. This also includes the case of reducing an $n$-particle state to its $r$-body reduced density matrix (as its marginal), also known as the $r$-body $N$-representability problem~\cite{klyachko2004quantum,schilling2014quantum}.

\subsection{Constrained symplectic transformations}\label{sec:constrained}
We now focus on bosonic systems with $N$ modes characterized by $2N$ quadrature operators $\hat{\xi}^a$ and their commutations relations, given by
\begin{equation}\label{commutation.rel}
    \hat{\xi}=(\hat{q}_1,\hat{p}_1,\dots,\hat{q}_N,\hat{p}_N)\,,\quad [\hat{\xi}^a,\hat{\xi}^b]=\ii\Omega^{ab}=\ii\bigoplus^N_{i=1}\left(\begin{array}{cc}
        0 & 1\\
        -1 & 0
    \end{array}\right)\,.
\end{equation}
Our Hilbert space $\mathcal{H}$ is naturally of the form~\eref{eq:Hspace} where each factor $\mathcal{H}_i$ represents a copy of $L^2(\mathbb{R})$ with a natural action of $\hat{\xi}_i=(\hat{q}_i,\hat{p}_i)$ as position and momentum operators. For each normalized quantum state $\ket{\psi}$ with $\braket{\psi|\hat{\xi}^a|\psi}$, \ie with vanishing quadrature expectation values, we can compute its covariance matrix
%\footnote{Let note that some authors use the alternative convention $C^{ab}=\frac{1}{2}\braket{\psi|\hat{\xi}^a\hat{\xi}^b+\hat{\xi}^b\hat{\xi}^a|\psi}$. \lfh{Mario may want to add a comment that this definition is more natural when calling it $C$ covariance matrix.}} 
as
\begin{equation}
    C^{ab}=\braket{\psi|\hat{\xi}^a\hat{\xi}^b+\hat{\xi}^b\hat{\xi}^a|\psi}\,\label{eq:covariance-matrix}
\end{equation}
A \emph{pure Gaussian state} is such a quantum state $\ket{\psi}$, for which we have $(C\Omega)^2=-\id$, which is equivalent to the property that the covariance matrix $C$ can be written as
\begin{equation}
    C=SS^\intercal\,,
\end{equation}
where $S$ must be a symplectic transformation satisfying
\begin{equation}
    S\Omega S^\intercal=\Omega\,.\label{eq:S-definition}
\end{equation}
Note, however, that this mapping is not one-to-one. While each symplectic matrix $S$ fixes $C$ uniquely, two different symplectic matrices $S$ and $\tilde{S}$ will give rise to the same covariance matrix $C$ if and only if $\tilde{S}=Su$, where $u$ is a symplectic matrix satisfying $uu^\intercal=\id$.

The restriction of a pure Gaussian state to individual modes yields in general mixed Gaussian states that are fully characterized by the restrictions of the covariance matrix to the respective subsystem, \ie
\begin{equation}
    \rho_i=\mathrm{Tr}_{\mathcal{H}_i^c}\ket{\psi}\bra{\psi}=\frac{\exp(-\hat{\xi}_iq\hat{\xi}_i)}{\Tr(\exp(-\hat{\xi}_iq_i\hat{\xi}_i))}\,,
\end{equation}
where the $2$-by-$2$ matrix $q$ can be computed from the restricted covariance matrix %\lfh{consider to include the exact formula to express $q_i$ in terms of $C_i$. DONE - see below.}
\begin{equation}\label{covariance}
C_i^{ab}=\braket{\psi|\hat{\xi}^a_i\hat{\xi}^b_i+\hat{\xi}^b_i\hat{\xi}^a_i|\psi}\,,
\end{equation}
which is the respective restriction of $C$ to the $i$-th $2$-by-$2$ diagonal block. The explicit relation between $q_i$ and $C_i$ is given by~\cite{hackl2021bosonic}
\begin{equation}
    q_i=\Omega_i^{-1} \mathrm{arccot}(-C_i\Omega_i^{-1})\,,
\end{equation}
where $\Omega_i$ refers to the symplectic form in the subsystem $i$, as indicated in~\eref{commutation.rel}.

For later considerations, it will be useful to consider two distinct scenarios:
\begin{itemize}
    \item \textbf{Scenario I: Arbitrary correlations between subsystem modes.} This was the scenario also considered in~\cite{5authors}, where we considered $n$ bosonic modes representing individual subsystems $i$ constrained to have the marginals $\rho_i$ characterized by $2$-by-$2$ subsystem covariance matrices $C_i$. Here, we define $\widehat{S}$ to be the $2N$-by-$2N$ matrix consisting only of the diagonal $2$-by-$2$ blocks of $SS^\intercal$, \ie
    \begin{equation}
        \widehat{S}=\mathrm{diag}([SS^\intercal]_1,\dots,[SS^\intercal]_n)\,,\label{eq:Shat}
    \end{equation}
    where $[SS^\intercal]_i=([SS^\intercal]_{kl})^{2i}_{k,l=2i-1}$. Similarly, we define the constraint matrix
    \begin{equation}
        \widehat{C}=\mathrm{diag}(C_1,\dots,C_N)\,,\label{eq:Chat}
    \end{equation}
    where $C_i$ are the $2$-by-$2$ covariance matrices of the individual subsystems. The constraint on a symplectic matrix $S$ then simply reads
    \begin{equation}
        \widehat{S}=\widehat{C}\,,\label{eq:constraint}
    \end{equation}
    as our ``widehat'' operation projects $SS^\intercal$ only on the diagonal blocks, which we want to constrain.
    \item \textbf{Scenario II: Vanishing correlations within certain sets of subsystem modes.} We consider $n$ collections of $N_j$ (with $j=1,\dots,n$) bosonic modes each, so that we have $N=\sum^{n}_{j=1}N_j$ modes in total. We do not only require to have marginals $\rho_{(j,i)}$ characterized by $2$-by-$2$ subsystem covariance matrices $C_{(j,i)}$ with $i=1,\dots,N_j$, but furthermore do not allow for any correlations between modes sharing the same $j$, \ie modes within the same collection.
    For this, we modify the definition $[SS]_j$ to be the $j$-th diagonal block with dimensions $2N_j$-by-$2N_j$. Similarly, we define $C_{j}=\mathrm{diag}(C_{(j,1)},\dots,C_{(j,n_j)})$ as the respective $2N_j$-by-$2N_j$ constraint matrix. Using these $[SS^\intercal]_j$ and $C_j$ in~\eref{eq:Shat} and~\eref{eq:Chat} will allow us to express the constraint again as~\eref{eq:constraint}, but now $S$ is enforced to now yield any correlations between $C_{(j,i_1)}$ and $C_{(j,i_2)}$.
\end{itemize}
The motivation for the second scenario is the case of Hawking radiation, where it is expected that there are little or no correlations between Hawking modes within the same epoch, as discussed in~\cite{Wald75,wald1994quantum}. However, the same argument can also be made for a general isolated quantum systems, where we may encounter physical systems with certain interactions and symmetries that imply that certain modes will never become correlated.

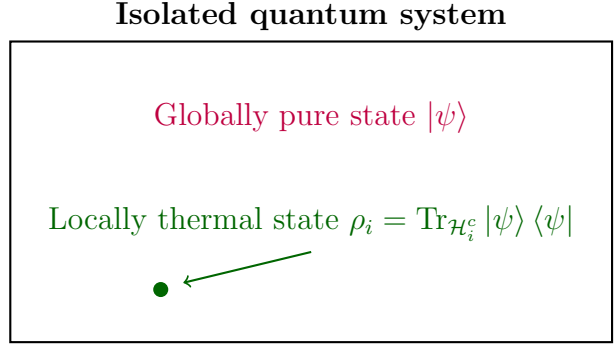
\begin{figure}
    \centering
    \begin{tikzpicture}
        \draw[thick] (6,-4) rectangle (14,-8);
			\draw (10,-4) node[above]{\textbf{Isolated quantum system}};
			\draw[purple] (10,-5) node{Globally pure state $\ket{\psi}$};
			\begin{scope}[yshift=-3mm]
				\draw[dgreen,<-,thick] (8.3,-6.91) -- (10,-6.5) node[above]{Locally thermal state $\rho_i=\Tr_{\mathcal{H}_i^c}\ket{\psi}\bra{\psi}$};
				\fill[dgreen] (8,-7) circle(1mm);
			\end{scope}
    \end{tikzpicture}
    \caption{Illustration of the simple idea that a globally pure state $\ket{\psi}$ may look thermal in subsystems $i$. Specifically, we consider $\ket{\psi}$ with fixed marginals $\rho_i=\Tr_{\mathcal{H}_i^c}\ket{\psi}\bra{\psi}$ that are thermal.}
    \label{fig:enter-label}
\end{figure}

We now want to construct the ensemble of all symplectic transformations (and thus global covariance matrices $C=SS^\intercal$) subject to the constraint $\widehat{S}=\widehat{C}$. While the symplectic transformations $S$ as a whole form a group, the ensemble of such constrained symplectic transformations subject to~\eref{eq:constraint} do not form a group. Nonetheless, they form a natural and in fact compact sub manifold of the symplectic group. We can therefore restrict the natural invariant volume form of the symplectic group to this sub manifold to evaluate ensemble averages. Given a general function $f(C)$ with $C=SS^\intercal$, we can evaluate its ensemble average as
\begin{equation}\label{eq:ensemble-average}
\braket{f(C)}=\frac{\int_{\mathbb{R}^{2N\times 2N}}f(C)\delta(\widehat{C}-\widehat{S})\delta(\Omega-S\Omega S^\intercal){\rm d}S}{\int_{\mathbb{R}^{2N\times 2N}}\delta(\widehat{C}-\widehat{S})\delta(\Omega-S\Omega S^\intercal){\rm d}S}\,,
\end{equation}
where the delta distributions apply to the independent components of the respective matrix arguments. For $\delta(\widehat{C}-\widehat{S})$ those are the symmetric parts of the diagonal blocks, hence of order $N$ in number.
For $\delta(\Omega-S\Omega S^\intercal)$
they are of strictly upper triangular
matrix entries of the antisymmetric matrix in the argument, the components of which are of order $N^2$ in number.
The volume form ${\rm d}S$ is the standard volume form in $\mathbb{R}^{2N\times 2N}$.
It can be checked that if all diagonal blocks
are constrained, as in the setting considered here, the resulting volume is finite, even though the
volume of
$\mathbb{R}^{2N\times 2N}$ of course is infinite.

This ensemble was constructed in~\cite{5authors} for Scenario I, which was then used to study typical correlations between two modes within the ensemble. We will also consider Scenario II noticing that the structure of the respective calculation and in particular the required saddle point approximation is the same. However, while~\cite{5authors} focused on the correlations between just two chose modes, we will study $f(C)$ to represent the von Neumann entanglement entropy $S_A(\ket{\psi})$ across a small subsystem $A$ and its complement.

\section{Typical reduced state in a small subsystem}
We now study what the typical reduced state $\rho_{C_A}$ of a small system looks like when we draw a random pure $\rho$ within Gaussian pure state ensemble with fixed marginals introduced in section~\ref{sec:Gaussian-fixed-marginals}.

\subsection{Reduced state of a subsystem}
Let $A$ be a subset of $N_A$ modes and $\rho_{C_A}$ the reduced density matrix on ${\cal H}_A$, the corresponding Hilbert space~\footnote{This Hilbert space is $\otimes_{i=1}^{N_A} {\cal H}_i^{\mathrm{osc}}$ where 
${\cal H}_i^{\mathrm{osc}}$
is the Hilbert space of the
$i$'th oscillator in $A$.
Formally ${\cal H}_A$
and each ${\cal H}_i^{\mathrm{osc}}$ is infinite-dimensional, but only states for which
$n_i$ is not much more than $\frac{\hbar\omega_i}{k_B T_i}$ are excited;
$\omega_i$ and $T_i$ are here the frequency and Hawking temperature pertaining to mode $i$.
}.
For Gaussian states, compare~\eref{eq:covariance-matrix}, we have
\begin{equation}
\label{eq:definition-Gaussian-rho-A}
\rho_{C_A}
= \Tr_{{\cal H}_B}
\left[\rho_C\right]
\,:\quad
\Tr\left[\rho_{A} \mathcal{D}_{{\rr}} \right] = 
e^{-\frac{1}{4}
{\rr}^\intercal\Omega^\intercal  C_A \Omega\, {\rr}}\,,
\end{equation}
where $B$ is the complement of $A$ and $C_A$ is the restriction of the full correlation matrix $C$ to the modes in $A$. We may hence denote the reduced density matrix $\rho_{C_A}$.
Among all states on ${\cal H}_A$ having correlations
$\widehat{C}_A$, the constraint
matrix 
$\widehat{C}$ restricted to $A$, the one of maximum (von Neumann) entropy is the 
product state
$\otimes_{i=1}^{N_A}\rho_{\widehat{C}_{i}}$
\footnote{Since Wald's off-diagonal constraints are that correlations between different modes in the same time window vanish they do not contribute to the maximum-entropy distribution which is built on the one-mode (thermal) constraints only.}. Writing $\widehat{C}_{i}=\id_{2\times 2}\lambda_i$ we have for any positive integer power $x$ \cite{Serafini}
\begin{equation}
\label{eq:trace-rhoCii-power-x}
\Tr\left[
\left({\rho}_{\widehat{C}_{i}}\right)^x\right]
= \frac{2^x}{(\lambda_i+1)^x-(\lambda_i-1)^x}
\end{equation}
In particular, the \textit{purity} is
\begin{equation}
\label{eq:purity-rhoCii}
\Tr\left[
\left({\rho}_{\widehat{C}_{i}}\right)^2\right]
= \frac{1}{\lambda_i}
\end{equation}
and the \textit{entropy} is, by differentiating and analytical continuation in $x$, 
\begin{eqnarray}
\label{eq:entropy-rhoCii}
\fl\Tr\left[-
{\rho}_{\widehat{C}_{i}} \log {\rho}_{\widehat{C}_{i}} \right]= -\log 2 + \frac{1}{2}
(\lambda_i+1)\log(\lambda_i+1)
- \frac{1}{2} (\lambda_i-1)\log(\lambda_i-1)
\end{eqnarray}
The best possible match in $A$
of a pure Gaussian state on all modes to
the constraints imposed by Hawking radiation, using no other information, is that it would reproduce 
the product state
$\otimes_{i=1}^{N_A}\rho_{\widehat{C}_{i}}$ and
\eref{eq:trace-rhoCii-power-x}.
We
will see that this is indeed the case, provided that $N_A$ is much smaller than ${{N}}$. The leading corrections go, as we will show, as
$\frac{N_A}{{N}}$. We start with the purity
\eref{eq:purity-rhoCii} which is technically simpler to obtain than 
the general result 
\eref{eq:trace-rhoCii-power-x},
and then
\eref{eq:entropy-rhoCii}.

\subsection{Average purity of a small subsystem}
The inverse relation to~\eref{eq:definition-Gaussian-rho-A} reads
\begin{equation}
\label{eq:definition-Gaussian-rho-A-inverse}
\rho_{C_A}
=
\frac{1}{(2\pi)^{N_A}}
\int d^{2N_A} 
{\rr_A}\,
\chi_{C_A}(\rr_A)
\mathcal{D}_{-\rr_A}
\end{equation}
where
$\chi_{C_A}(\rr_A) \equiv
e^{-\frac{1}{4}
{\rr_A}^\intercal\Omega^\intercal  C_A \Omega\, {\rr_A}}$.
From this follows
\begin{equation}
\label{eq:purity-rhoA}
\Tr\left[
\left({\rho}_{C_A}\right)^2\right]
= 
\int
\frac{d^{2N_A} 
{\rr_A}}{(2\pi)^{N_A}}\,
\chi_{C_A}^2(\rr_A)
=\frac{1}{\det\left(C_A\right)^{\frac{1}{2}}}
\end{equation}
For a product state this is the same 
as \eref{eq:purity-rhoCii}.
Writing now 
$C_A=\Pi_A SS^\intercal  \Pi_A^\intercal$, the projection of a total pure state correlation matrix to the modes in $A$, 
$\widehat{S}$
for the elements of the total correlation matrix which appear in the constraint $\widehat{C}$,
and
$d\mu_{\widehat{C}}(S)
=
dS\,
\delta\left(\widehat{S}-\widehat{C}\right)
\delta\left(
S\Omega S^\intercal-\Omega\right)$
for a measure on 
real symmetric matrices $S$ which enforce that they are symplectic
and match the constraints
we have
\begin{equation}
\label{eq:average-purity-1}
\left\langle\Tr\rho_{C_A}^2\right\rangle=
\frac{
\int d^{2N_A} 
{\rr}\,
\int d\mu_{\widehat{C}}S\, e^{-\frac{1}{2}
{\rr}^\intercal\Omega^\intercal  \Pi_A SS^\intercal\Pi_A^\intercal \Omega\, {\rr}}}
{(2\pi)^{N_A} \int d\mu_{\widehat{C}}S\,}
\end{equation}
Expressing the delta functions in the Fourier domain using two auxiliary multiplier matrices $A$ and $B$ this can be written
\begin{equation}
\label{eq:average-purity-2}
\left\langle\Tr\rho_{C_A}^2\right\rangle=
\frac{
\int
d^{2N_A} 
{\rr}\,
d\nu_{AB}\, 
\det[(\ii A+{\rr}
{\rr}^\intercal)\otimes\id_2+B\otimes\tau_2
]^{-\frac{{{N}}}{2}}}
{
(2\pi)^{N_A}
\int d\nu_{AB}\, 
\det[\ii A\otimes\id_2+B\otimes\tau_2]^{-\frac{{{N}}}{2}}} \nonumber
\end{equation}
where $d\nu_{A,B}$
is an integration measure over $A$ and $B$ including appropriate convergence factors,
$\id_2$ is the identity matrix on one mode
and $\tau_2=\left(\begin{array}{cc}0 & -\ii \\ \ii & 0
\end{array}\right)$.
For details of the construction and the representation using
$A$ and $B$,
see \cite{5authors} and Appendix. 
The two $2{{N}}$-dimensional 
determinants
in the numerator and denominator 
can be reduced to ${{N}}$-dimensional
determinants canceling $\frac{1}{2}$
in the exponent.
They are then
the same up to a rank-1 perturbation, and
we have
\begin{equation}
\det[\ii A+B +
{\rr}
{\rr}^\intercal
]=
\det[\ii A+B](1+{\rr}^\intercal\frac{1}{\ii A+B}{\rr})
\end{equation}
By a change of variable we hence have
\begin{equation}
\label{eq:average-purity-3}
\left\langle\Tr\rho_{C_A}^2\right\rangle=
\frac{
\int
d\nu_{AB}\, 
\det[\ii A+B]^{-{{N}}}\,
\det^{-\frac{1}{2}}\tilde{A}
}
{
C(N_A,{{N}})
\int d\nu_{AB}\, 
\det[\ii A+B]^{-{{N}}}
}
\end{equation}
where $\tilde{A}$
is the symmetric sum of
$(\ii A+B)^{-1}$ and its transpose and
\begin{equation}
    C(N_A,{{N}})
    = (2\pi)^{N_A}
\left(\int
d^{2N_A} 
{\rr} \, \frac{1}{\left(1+{\rr}^\intercal
{\rr}\right)^{{N}}} \right)^{-1}
\end{equation}
This integral can be evaluated by recursion and
estimated using Stirling's formula in the
limit when ${{N}}$ is much larger than 
$N_A$.
%\footnote{Erik only did this partly. It appears to him now that in the notes the final expression for $\int_0^{\infty} (1+x^2)^{-N} dx$ and a factor $({{N}})^{\frac{1}{2}}$ from the stationary phase solution are missing.}.
In \eref{eq:average-purity-3} a stationary phase analysis can be done using only the 
measure 
$d\nu_{AB}$ and the determinant 
$\det[\ii A+B]$ raised to a high power; the outcome are stationary points 
in $A$ and $B$ such that
$\tilde{A}=\frac{1}{{N}}\widehat{C}_A$.
Combing with $C(N_A,{{N}})$ and comparing to 
\eref{eq:purity-rhoA}
we have hence shown that
\begin{equation}
\label{eq:average-purity-4}
\left\langle\Tr\rho_{C_A}^2\right\rangle=
\Tr\rho_{\widehat{C}_A}^2
\end{equation}
The average purity on ${\cal H}_A$ of a total random pure Gaussian state satisfying the constraints $\widehat{C}$, inside and outside $A$, is therefore the same as for a mixed Gaussian
state on $A$ with correlation matrix 
$\widehat{C}_A$.
For details we refer again to Appendix. 

\subsection{Average higher moments of $\rho_{C_A}$}
To compute $\langle\Tr \rho^x_{C_A}\rangle$
for $x=3,4,\ldots$ we represent $x-1$ copies  
of $\rho$ using \eref{eq:definition-Gaussian-rho-A-inverse},
and take the trace with the last remaining copy.
Thus
\begin{eqnarray}
\label{eq:Tr-rhoA-x}
\Tr \rho^x_{C_A}
&=&
\Tr\Big[\rho_{C_A} \,
\frac{1}{(2\pi)^{N_A(x-1)}}\,
\int \prod_{r=1}^{x-1}\, d^{2N_A} 
{\rr}^r\, 
%\nonumber \\ && \qquad\qquad\qquad
\chi_{C_A}(\rr_A)
\mathcal{D}_{-\rr_A}\Big] 
\end{eqnarray}
Using the group property of coherent state operators we have
\begin{equation}
    \prod_{r=1}^{x-1}\,
\mathcal{D}_{-\rr_A}
= \mathcal{D}_{-\sum_{r=1}^{x-1}\rr_A}
{\cal F}\left(\{-\rr_A\}_{r=1}^{x-1}\right)
\end{equation}
where
\begin{equation}
{\cal F}\left(\{\rr_A\}_{r=1}^{x-1}\right) = 
e^{\frac{1}{2}\sum_{a=1}^{x-1}\sum_{b=r+1}^{x-1}\left(\alpha^{(a)}
(\alpha^{(b)})^* - (\alpha^{(a)})^*\alpha^{(b)}\right)} \nonumber
\end{equation}
Hence, we have the representation
\begin{eqnarray}
\label{eq:Tr-rhoA-x-2}
\fl\Tr \rho^x_{C_A}
&=
 \frac{1}{(2\pi)^{N_A(x-1)}}
\int \prod_{r=1}^{x-1}
\left[d^{2N_A} 
{\rr}^r\,
\chi_{C_A}(\rr_A)\right]
\chi_{C_A}\left(-\sum_{r=1}^{x-1}\rr_A\right)
{\cal F}\left(\{\rr_A\}_{r=1}^{x-1}\right)
\end{eqnarray}
We note that for $x=2$ (the previous case, purity), there is only one integration over coherent state parameters, and ${\cal F}=1$. It is convenient to group all the coherent state 
parameters
in \eref{eq:Tr-rhoA-x-2} as one vector variable $\mathbf{v}$.
Then, in analogy with \eref{eq:average-purity-2}, we have
\begin{eqnarray}
\label{eq:average-trace-rhoA-x}
\fl\left\langle\Tr\rho_{C_A}^x\right\rangle &=
\frac{1}
{
(2\pi)^{N_A(x-1)}
\int d\nu_{AB}\, 
\det[\ii A+B]^{-{{N}}}
}
\int \frac{d\nu_{AB} d\mathbf{v}\,{\cal F}\left(\mathbf{v}\right)} 
{
\det\left[\ii A+B +
\frac{1}{4}\mathbf{v} \cdot M
\mathbf{v}^\intercal
\right]^{{N}}}
\end{eqnarray}
where
\begin{equation}
    M = \id_{x-1} + ee^\intercal
    \qquad e=\left(1,1,\ldots,1\right)
\end{equation}
and where the notation $\mathbf{v} \cdot M\mathbf{v}^\intercal$ means a rank-1 matrix of
size ${{N}}$ \textit{i.e.} the scalar product goes over the $x-1$ "replica" dimensions.
Similarly to above
the ratio of the two determinants is
\begin{displaymath}
\det\left(\id_x +
\frac{1}{4} \left(\mathbf{v}^\intercal\cdot 
\frac{1}{\ii A+B} \mathbf{v}\right)
M\right)
\end{displaymath}
where now $\mathbf{v}^\intercal\cdot 
\frac{1}{\ii A+B} \mathbf{v}$
means a $x-1$-dimensional matrix.
The symmetric sum of
the stationary point of
$\frac{1}{\ii A+B}$ and its transpose
is $\frac{1}{{N}}\widehat{C}$.
Inserting this in the ratio of the two determinants to inverse power of ${{N}}$ and taking the limit 
of large ${{N}}$
we have 
\begin{displaymath}
e^{-\frac{1}{4}
\Tr
\left[\left(\widehat{C}\otimes M\right)
\mathbf{v}\mathbf{v}^\intercal\right]}
\end{displaymath}
Comparing to
\eref{eq:Tr-rhoA-x-2} we see that it is the same expression, which hence 
implies
\begin{eqnarray}
\label{eq:Tr-rho-power-x-3}
\left\langle\Tr_A
\left[\rho_{C_A}^x\right]\right\rangle
&= 
\Tr_A
\left[\rho_{\widehat{C}_A}^x\right]
\end{eqnarray}
The average trace of $\rho_{C_A}$ to the $x$'th power of a total random
pure Gaussian state satisfying the constraints $\widehat{C}$ is thus the same as for a mixed Gaussian state
on ${\cal H}_A$ with correlation matrix 
$\widehat{C}_A$. 
In Appendix we show that this also follows from an explicit  
calculation of the integral
in
\eref{eq:Tr-rhoA-x-2},
which for a factorized 
constraint matrix $\widehat{C}$ leads to
\eref{eq:trace-rhoCii-power-x}.

\subsection{Replica analysis and average entropy}
As the final result for the average
trace of the reduced density matrix
on ${\cal H}_A$
to power $x$,
the product of
\eref{eq:trace-rhoCii-power-x}
over the modes in $A$, 
is an explicit and clearly analytic function of $x$, we can 
analytically continue
to $x=1$ and take the derivative.
The average entropy
of the reduced density matrix
on ${\cal H}_A$
is therefore the sum of 
\eref{eq:entropy-rhoCii}
over the modes in $A$.
In words, the average entropy of 
$\rho_{C_A}$ is to leading
order as large as it can be 
given $\hat{C}_A$, the constraint
correlation matrix on the modes
in $A$, while the 
constraints on the correlation matrix
between modes inside and outside $A$
or wholly outside $A$ do not matter.

Let us emphasize that this finding will in general only apply to small subsystems, \ie in the asymptotic limit with $N\to\infty$ and $N_A/N\to 0$. In particular, it will not apply ``volume''-sized subsystem making up a finite fraction $N_A/N\to f$ in the large $N$ limit. This follows from the fact that the reduced states $\rho_{C_A}$ and its complementary reduction $\rho_B$ must have the same eigenvalues (apart from zero), while the associated $\lambda_i$ in either subsystems must only satisfy the much weaker inequalities derived in~\cite{Eisert2008}. It should therefore not surprise that for subsystems of comparable size the spectral properties of $\rho_{C_A}$ and $\rho_B$ must depend on both $\widehat{C}_A$ and $\widehat{C}_B$.\\

\section{Application to Hawking radiation}
So far, we derived a general simple general result for the ensemble introduced in~\eref{eq:ensemble-average}, namely that for $N_A/N\to 0$ the typical state $\rho_A$ in the subsystem $A$ is the Gaussian mixed state with covariance matrix $\widehat{C}_A$. We will now apply this result to the case of Hawking radiation.

\subsection{The random Gaussian pure state model of Hawking radiation}
Hawking's theory is based on modes of radiation escaping from an eternal black hole in finite time windows. The width in time of the windows does not matter; it is compensated by the frequency widths of the modes. When adapting this to an evaporating black hole it is convenient to choose the widths of the time windows as $\Delta t=t_P \left(\frac{M(t)}{m_P}\right)^2$ where $t_P$ is Planck time, $m_P$ is Planck mass, and $M(t)$ is the remaining mass of the black hole at time $t$. Up to a constant depending on the number of fields taken into account, $\Delta t$ is the time it takes a black hole of mass $M(t)$ to emit Hawking radiation of energy $m_P\, c^2$. In each time window, the highest excited frequency is about $\frac{k_B T_H(M)}{\hbar}$ where $T_H(M)$ is the Hawking temperature, hence (up to constants) $\frac{m_P c^2}{\hbar}\frac{m_P}{M}$. The frequency discretization in the same window is $\Delta\omega=\frac{1}{\Delta t}$
such that there are about $\frac{M(t)}{m_P}$ excited modes in the time window. Using that in each time window $\Delta t$ the black hole radiates away mass $k\, m_P$
where $k$ is a numerical constant,
the total number of excited modes in the lifetime of the black hole  is about
\begin{equation}
\label{eq:total-number-of-modes}
{{N}}_{M(0)}
=
\int_0^{M(0)}
\frac{m}{m_P}\frac{dm}{k\, m_P}
= \frac{1}{2k}\frac{M^2(0)}{m^2_P}\,,
\end{equation}
which for astrophysical black holes give the approximate value quoted above.
% A Gaussian state of ${{N}}_{M(0)}$
% bosonic modes has a density matrix $\rho_C$ satisfying
% \begin{equation}
% \label{eq:definition-Gaussian-rho}
% \Tr\left[\rho_C \mathcal{D}_{\alpha} \right] = 
% e^{-\frac{1}{4}
% {\rr}^\intercal\Omega^\intercal  C \Omega\, {\rr}}
%     \qquad  \mathcal{D}_{\alpha}
%     =
%     e^{\alpha a^{\dagger}-\alpha^* a}
% \end{equation}
% where $\mathcal{D}_{\alpha}$ is the coherent state operator, $a^{\dagger}$ and $a$ are the 
% creation/annihilation operators,
% $\Omega$ is the symplectic form on phase space,
% and $\alpha$ and $r$ are two ways to write the
% coherent state parameters.
% The Gaussian state is fully chracterized by the covariance matrix $C$, which can always be written
% $S\,\mathrm{diag}\left(d_1,d_1,\ldots,d_{{{N}}_{M(0)}},d_{{{N}}_{M(0)}}\right)\,S^\intercal$
% where the $d_i$s are the
% symplectic eigenvalues and $S$ is a 
% $2{{{N}}_{M(0)}}\times 2{{{N}}_{M(0)}}$-dimensional symplectic matrix.
% For a Gaussian pure state all the symplectic eigenvalues are equal to one, and
% $C=SS^\intercal$.

The predictions of Hawking's theory are in the language of Gaussian states that the single-mode correlation matrices are thermal
$\id_{2\times 2}\,\exp(\frac{j\hbar \Delta w_{t}}{k_B T_t})$ where $T_t$ and 
$\Delta w_{t}$ are the temperature and frequency discretization at time $t$ and $j=1,2,\ldots$ counts the excited modes in that time window.
According to the results of
Wald mode-mode correlations in the same time window are zero. All these data can be collected into a constraint correlation matrix $\widehat{C}$ which has a block structure.

It was shown in~\cite{Aurell2022} that there are many pure Gaussian states that can satisfy these constraints, \ie that looks thermal with temperature $T$ when considering individual modes only. This follows from mathematical inequalities derived in continuous-variable quantum information theory \cite{Eisert2008}. One can thus study the distribution of mode-mode correlations not constrained by $\widehat{C}$, and from that the probability that two such modes are entangled. As shown in~\cite{5authors}, most such mode-mode correlations are small and with high probability no such pair of modes is entangled in the random pure Gaussian state ensemble.

\subsection{The Page curve and fluctuations}
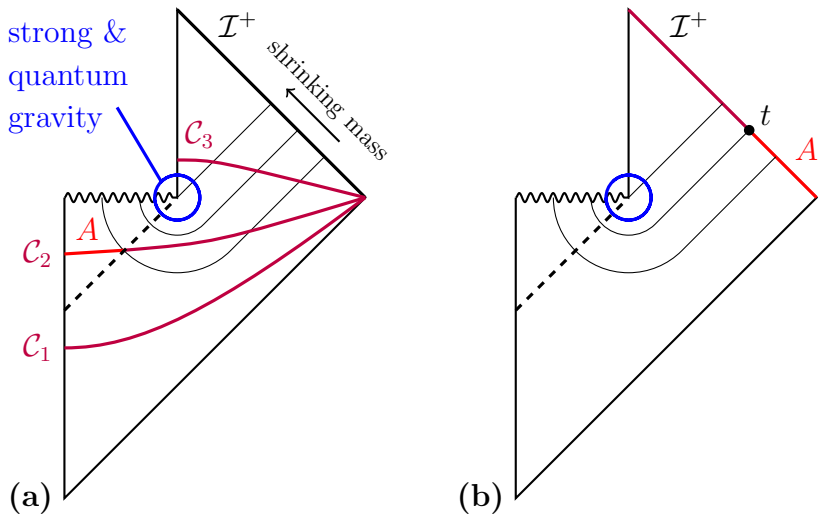
\begin{figure}
    \centering
    \begin{tikzpicture}
        \begin{scope}
			\draw[thick] (0,0) -- (0,-4) node[left]{\textbf{(a)}} -- (4,0) -- (1.5,2.5) -- (1.5,0);
			\draw[very thick] (4,0) -- (1.5,2.5) node[right,yshift=-2mm,xshift=4mm]{$\mathcal{I}^+$};
			\draw[thick, snake it] (0,0) -- (1.5,0);
			\draw (1.5,0) -- (2.75,1.25);
			\draw (1,0) arc (-180:-45:.5) --(3.10355,0.896447);
			\draw (.5,0) arc (-180:-45:1) --(3.45711,0.542893) ;
                \begin{scope}[xshift=2mm,yshift=2mm]
                 \draw[thick,->] (3.45711,0.542893) -- node[xshift=2mm,yshift=2mm,rotate=-45,scale=.8]{shrinking mass} (2.75,1.25);
                \end{scope}

			\draw[very thick,blue] (1.2,0) arc (-540:135:.3) -- (.7,1.2) node[above,text width=1.7cm,align=left,xshift=-.6cm,yshift=-.5cm]{strong \& \\
				quantum\\gravity};
                \draw[very thick,purple] (4,0) .. controls (2,-1.5) and (1,-2) .. (0,-2) node[left]{${\cal C}_1$};
                \draw[very thick,purple] (4,0) .. controls (2,-.6) .. (.8,-.7);
                \draw[very thick,red] (.8,-.7) -- node[above,xshift=-1mm]{$A$} (0,-.75) node[left,purple]{${\cal C}_2$};
                \draw[very thick,purple] (4,0) .. controls (2,.5).. (1.5,.5) node[above,xshift=3mm]{${\cal C}_3$};
			\draw[very thick, dashed] (0,-1.5) -- (1.5,0);
		\end{scope}
        \begin{scope}[xshift=6cm]
			\draw[thick] (0,0) -- (0,-4) node[left]{\textbf{(b)}} -- (4,0) -- (1.5,2.5) -- (1.5,0);
                \draw[very thick,red] (4,0) -- node[right,yshift=2mm]{$A$} (3.10355,0.896447);
                
			\draw[very thick,purple]  (3.10355,0.896447) -- (1.5,2.5) node[right,yshift=-2mm,xshift=4mm,black]{$\mathcal{I}^+$};
			\draw[thick, snake it] (0,0) -- (1.5,0);
			\draw[very thick, dashed] (0,-1.5) -- (1.5,0);
			\draw (1.5,0) -- (2.75,1.25);
            \draw[very thick,blue] (1.2,0) arc (-540:135:.3);
			\draw (1,0) arc (-180:-45:.5) --(3.10355,0.896447);
			\draw (.5,0) arc (-180:-45:1) --(3.45711,0.542893);
            \fill (3.10355,0.896447) node[right,yshift=2mm]{$t$} circle(2pt);
			%\fill[dgreen] (3.45711,0.542893) node[right,yshift=1mm]{$\rho_i$} circle(1mm);
			%\draw[very thick,blue] (1.2,0) arc (-540:135:.3) -- (.7,1.2) node[above,text width=1.5cm,align=left,xshift=-.5cm,yshift=-.4cm]{strong \& \\quantum\\gravity};
			
		\end{scope}
    \end{tikzpicture}
    \caption{This sketch shows the Penrose of an evaporating black hole based on~\cite{wald1994quantum}. We show two different ways to define a Page curve: \textbf{(a)} The entanglement entropy across spatial Cauchy slices $\mathcal{C}_i$ (purple lines) evolving with a time parameter $t$ and divided by the black hole event horizon (dashed line) with subsystem $A$ (indicated in red), see also~\cite{almheiri2021entropy}. \textbf{(b)} The entanglement entropy across null infinity $\mathcal{I}^+$ (purple line) with respect to a bi-partition characterized by a conformal parameter $t$ with subsystem $A$ (indicated in red). The entropy as a function of $t$ yields the \emph{Page curve of Hawking radiation}.}
    \label{fig:penrose-diagram-page}
\end{figure}

The standard Page curve as introduced by Don Page in~\cite{page1993average,Page1993} refers to the entanglement entropy between the inside and the outside of an evaporating black hole as a function of time. For this, one typically considers different Cauchy slices, as indicated in figure~\ref{fig:penrose-diagram-page}(a). However, the concept of the Page curve has been applied more broadly when considering an ensemble of pure quantum states in a large system with respect to a bi-partition. In the following, we apply this second more general concept of a Page curve to the quantum system describing Hawking radiation on future null infinity $\mathcal{I}^+$, as also illustrated in figure~\ref{fig:penrose-diagram-page}(b). We refer to this curve as the \emph{Page curve of Hawking radiation} as it does not describe the entanglement between the quantum field inside and outside the black hole, but rather the entanglement between different radiation modes at null infinity.

Therefore, we consider the Page curve as the average entanglement entropy $\braket{S_{A_t}(\ket{\psi})}$ as a function of conformal time $t$, where $A_t$ refers to all modes up to $t$. For a black hole state of initially low entropy, assimilated to a pure state, this should start out at zero, increase to a maximum at $t$ about half the life-time of the black hole, and then decrease to zero. 
Using \eref{eq:total-number-of-modes}, we have $N_{A_t}={{N}}_{M(0)}-{{N}}_{M(t)}=\frac{1}{2k}\frac{M^2(0)-M^2(t)}{m^2_P}$ and our results apply when $M(0)-M(t) \ll M(0)$, or when $t$ is much smaller than the life time of the black hole.
Our theory also gives the entropy of $N_A$ modes in time windows somewhere in the middle of the life of the black hole, but such estimates are not encoded by the Page curve.
Formally our results would apply to the end of the life-time of the black hole, but then the validity of Hawking's predictions are not certain, as the curvature around the black hole horizon eventually becomes large. Hence our theory only gives a small part of the Page curve, specifically its slope at the initial time.
  
Nevertheless it is of interest to outline how our analysis would have to be modified if $N_A$ would grow to be comparable to ${{N}}$. One type of corrections are the various pre-factors (normalizations) which above have only been estimated in the limit when ${{N}}$ tends to infinity. A second difference is that it would then not be correct to estimate the ratios of determinants as in 
\eref{eq:average-purity-3}
and
\eref{eq:average-trace-rhoA-x}.
Taking $A^*$ and $B^*$ the stationary points of
the expression in the denominator
(derived in detail in Appendix)
we should write $A=A^*+\Delta A$ and $B=B^*+\Delta B$ and expand also the rank-1 perturbation
correction term which depends on $\frac{1}{\ii A+B}$
in $\Delta A$ and $\Delta B$.
Such an analysis appears in general non-trivial, and have to be left to future work.
For the case of purity \eref{eq:average-purity-3} it is outlined to second (Gaussian) order in the fluctuations in Appendix; note however that this is not necessarily the only relevant order, and that this answer outlined in Appendix may be incomplete.
In any case, it is to be expected that these fluctuations around the stationary points also give corrections of relative size $\frac{N_A}{{N}}$, hence matching and/or competing with the variations of the pre-factors (normalizations).

\begin{figure}
    \centering
    \begin{tikzpicture}
        \draw[very thick,->] (0,0) -- (8,0) node[below]{$N_A/N$};
        \draw[very thick] (7,.1) -- (7,-.1) node[below]{$1$};
        \draw[very thick,->] (0,-.1) node[below]{0} -- (0,4.5) node[left]{$\braket{S_A}$};
        \draw[very thick, purple, dashed] (0,0) .. controls (3,3) and (5,4) .. (7,0);
        \draw[purple] (4,3) node{?};
        \draw[very thick,->,red] (0,0) -- (3,3);
        \draw[very thick,->,blue] (7,0) -- (5,4);
        
    \end{tikzpicture}
    \caption{We sketch what information about the Page curve can be extracted using our techniques. Specifically, we can get the initial and final slope corresponding, indicated by the red and blue arrow, while computing the exact intermediate curve requires methods to evaluate the average $\braket{S_A}$ for $0< N_A/N<1$. Let us highlight that--in contrast to scenarios typically studied in quantum many-body systems~\cite{bianchi2022volume}--our ensemble~\eref{eq:ensemble-average} is not invariant under subsystem permutations and thus generally also not symmetric.}
    \label{fig:page-curve}
\end{figure}
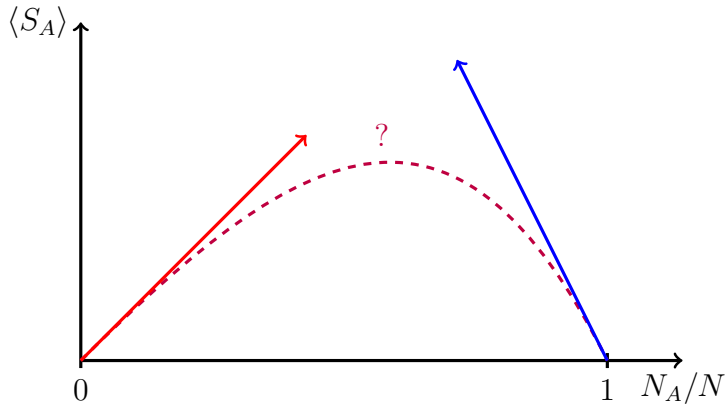

\section{Discussion}

\subsection{Relevance to the black hole information paradox}
There exist an enormous literature trying to determine how Hawking radiation actually is entangled, reviewed up to about a decade ago in \cite{Harlow,Marolf} and  with a string of high-profile contributions since
\cite{Almheiri2019,Almheiri2020,Penington2020,Almeiri21,Penington2022,Balasubramanian2024b,Balasubramanian2024a}.
As a branch of Physics the field of quantum black holes suffers from a lack of experimental data, as Hawking radiation has not been observed, and likely will never be observed directly. Therefore all predictions, be they ever so solid and satisfying theoretically, lack the robustness of conclusions made on hypotheses that have had to face the risk of being falsified by experiments.  

In this context the value of our approach is that it is based on relatively simple mathematics. It may be that, in fact, Hawking radiation from a black hole is in a total pure state which is not Gaussian. Of this we can say nothing. However, since the one-mode marginals are Gaussian in Hawking's theory, the total state would then have additional structure, and without a physical argument why this is so we are allowed to assume that the total state in Gaussian.
We can hence say that the black hole information paradox only pertains if one were able to do quantum tomography of states of astronomically high dimension, simply because the number of  modes  excited in Hawking radiation is so enormously large. If all one can do is to characterize states of smaller subsets of Hawking radiation, they would almost surely be indistinguishable from thermal radiation, if the total state would be picked randomly from the constrained Gaussian ensemble studied here.

\subsection{Justification of Gaussianity assumption}
As alluded to previously, crucial assumption in our symplectic ensemble is that the final state of radiation is a Gaussian state. This is well-justified for early and intermediate radiation, as tracing back into the causal past of these modes only covers regions of spacetime far away from the black hole horizon at sufficiently large mass, which implies that semi-classical approximation of a free scalar field propagating on a curved background is justified. Only the very late radiation, whose causal past traces back through a near-horizon region of sufficiently small black hole mass, may be highly non-Gaussian due to the large curvature (strong gravity) and potential quantum gravity effects.

An important question is therefore to what extent potential late time non-Gaussianities will affect the result of the present manuscript which were derived based on the Gaussian assumption. There certainly exist carefully constructed non-Gaussian states whose entanglement structure between early and very late radiation will by construction deviate significantly from our findings. However, we have good reasons to believe that extending our model to include some non-Gaussianities in the late radiation (and in its correlations with early radiation) will not affect the average entanglement entropy for small subsystems. This is due to the fact that the bi-partite entanglement entropy for random pure Gaussian states and general random pure states are generally the same. This was discussed in~\cite{bianchi2022volume} for fermionic systems, but can also be seen by comparing~\cite{youm2024average} for random pure Gaussian states with~\cite{yauk2024typical} for random pure states in bosonic systems. In short, the initial slope of the Page curve\footnote{Let us emphasize that we are referring to the Page curve along $\mathcal{I}^+$ as 
illustrated in Fig.~\ref{fig:penrose-diagram-page} describing the correlations between different radiation subsystems and \emph{not} the Page curve describing entanglement between the quantum field inside and outside of the black hole horizon.} as shown in figure~\ref{fig:page-curve} is the same for Gaussian and non-Gaussian random states.

\subsection{Outlook}
%\lfh{Discuss next order correction and how the result can certainly not apply for larger subsystems, as it is clearly incompatible with the symmetry of the entanglement entropy etc.}
%\lfh{Discuss two aspects: (1) Average entanglement entropy for small subsystems generally behaves the same for Gaussian and general non-Gaussian states. (2) Small subsystems of early modes are far away from any .}
The previous discussion has been limited to subsystems of small dimension ($N_A$), compared to the total dimension of the system ($N$). For the application to Hawking radiation it therefore only pertains to the beginning and the end of the Page curve (in the random Gaussian pure state ensemble model), but not to the arguably even more interesting intermediate Page time,
see figure~\ref{fig:page-curve}. This limitation can be traced to the mean-field approximation explicitly introduced after \eref{eq:average-purity-3}, and used again to derive \eref{eq:Tr-rho-power-x-3}. When the dimension of the subsystem is comparable to the total dimension fluctuations cannot be ignored, as well as systematic corrections from pre-factors which are of relative size $N_A/N$. In~\cite{5authors}, effects of fluctuations were evaluated to quadratic order, as this was necessary to arrive at a nontrivial result on mode-mode correlations. In the more complex setting of subsystem entropy, where there is also the
complicating influence of the
replica dimension $x$, we have to leave this for future work. We note however that as predictions on the full Page curve
from specific microscopic models
have been published in the recent literature
\textit{cf.} \cite{Almheiri2020b,Gautason2020,Marolf2021,Gan2022,Akal2022}, this would allow comparison on how those predictions agree, or perhaps differ from,  
the simpler random model investigated here. This would then in turn speak to if 
Hawking radiation as predicted by those models is, or is not, well described by a random Gaussian states.

From a general point of view our work gives further grounds to conjecture that Hawking's information paradox may well mostly be a consequence of geometry in high dimensions, as most of the paradoxical qualitative observations (small subsystem thermality, absence of mode-mode correlations) can be reproduced in a fairly simple random model.

\ack
We thank Robert Jonsson and Pawe{\l} Horodecki for inspiring discussions on the topic.
EA acknowledges support of the Swedish Research Council through grant 2020-04980.
LH acknowledges support by grant $\#$63132 and by grant $\#$62312, ‘The Quantum Information Structure of Spacetime’ Project (QISS), from the John Templeton Foundation and an Australian Research Council Australian Discovery Early Career Researcher Award (DECRA) DE230100829 funded by the Australian Government.
MK acknowledges support by the Australian Research Council via Discovery Project grant DP210102887. The opinions expressed in this publication are those of the authors and do not necessarily reflect the views of the respective funding organization.

\appendix

\section{Average purity of a small subsystem}
\label{sec:everage-purity}
We consider the more general Scenario II discussed in section~\ref{sec:constrained} for the case of Hawking radiation. We have $n$ time slices of $N_1,\ldots,N_N$ bosonic modes. Thus, the total number of modes is $N=\sum_{j=1}^nN_j$. In each time slice we fix the covariance matrices of the marginals to be $C_{(j)}=\mathrm{diag}(C_{(j,1)},\ldots,C_{(j,N_j)})$ which can be collected in one matrix $\widehat{C}=\mathrm{diag}(C_{(1)},\ldots,C_{(N)})\in\mathbb{R}^{2N\times 2N}$. We underline that in this matrix we do not only enter Hawking's prediction of thermal radiation via the $2\times 2$ matrices $C_{(j,i)}$ but also include the zero correlations of modes in the same time slice predicted by Wald~\cite{Wald75,wald1994quantum}.

We introduce with $\{SS^\intercal\}_{(j)}$ the $2N_j\times 2N_j$ at the $j$th position on along the diagonal of the matrix $SS^\intercal$ and $\widehat{S}=\mathrm{diag}(\{SS^\intercal\}_{(1)},\ldots,\{SS^\intercal\}_{(n)})\in\mathbb{R}^{2N\times 2N}$ Additionally, $\Pi_A$ shall be the orthogonal projection on the $2N_A$ dimensional subspace for the covariance matrix on the subsystem $A$. Then, the expected purity of system $A$ is
\begin{equation}\label{purity.start}
\left\langle\Tr\rho_{C_A}^2\right\rangle=\frac{\int_{\mathbb{R}^{2N\times 2N}}\det(\Pi_ASS^\intercal\Pi_A^\intercal)^{-1/2}\delta(\widehat{C}-\widehat{S})\delta(\Omega -S\Omega S^\intercal){\rm d}S}{\int_{\mathbb{R}^{2N\times 2N}}\delta(\widehat{C}-\widehat{S})\delta(\Omega -S\Omega S^\intercal){\rm d}S}\,.
\end{equation}
For calculating this average we follow the same ideas as in our previous work with Robert Jonsson and
Pawe{\l} Horodecki~\cite{5authors}.

First of all we introduce a real antisymmetric $2N\times 2N$ matrix $B\in{\rm ASym}(2N)$ to write the Dirac delta function enforcing the symplectic structure of $S$ and a block-diagonal matrix $\widehat{A}={\rm diag}(A_{(1)},\ldots,A_{(N)})$ with real symmetric matrices $A_{(j)}\in{\rm Sym}(2N_j)$ which represent the Dirac delta function for the marginals. Moreover, we need to introduce a real Gaussian random vector $r\in\mathbb{R}^{2N_A}$ which is embedded in $\mathbb{R}^{2N}$ by $\rr_A=\Pi_A^\intercal \rr$. Then, it is
\begin{eqnarray}
\fl&\left\langle\Tr\rho_{C_A}^2\right\rangle=\lim_{\tilde{\epsilon}\to0}\frac{\int {\rm e}^{-r_A^\intercal SS^\intercal r_A+\Tr(\epsilon\id_{2N}-\ii A)(\widehat{C}-SS^\intercal)+\Tr(\Omega -S\Omega S^\intercal)B+\tilde{\epsilon}\Tr(\epsilon\id_{2N}-\ii A)^2+\tilde{\epsilon}\Tr B^2}{\rm d}S{\rm d}A{\rm d}B{\rm d}r}{\int {\rm e}^{-r_A^\intercal r_A+\Tr(\epsilon\id_{2N}-\ii A)(\widehat{C}-SS^\intercal)+\Tr(\Omega -S\Omega S^\intercal)B+\tilde{\epsilon}\Tr(\epsilon\id_{2N}-\ii A)^2+\tilde{\epsilon}\Tr B^2}{\rm d}S {\rm d}A{\rm d}B{\rm d}r}.\nonumber \\
\fl&\label{app.A2}
\end{eqnarray}
Once we carry out the integration over $S$ we find
\begin{eqnarray}
\fl\left\langle\Tr\rho_{C_A}^2\right\rangle&=\lim_{\tilde{\epsilon}\to0}\frac{\int {\rm e}^{\Tr(\epsilon\id_{2N}-\ii A)\widehat{C}}\det[(r_A r_A^\intercal+\epsilon\id_{2N}-\ii A)\otimes\id_{2}+B\otimes\tau_2]^{-N/2}\ {\rm d}P(A,B){\rm d}r}{\int {\rm e}^{-r_A^\intercal r_A+\Tr(\epsilon\id_{2N}-\ii A)\widehat{C}}\det[(\epsilon\id_{2N}-\ii A)\otimes\id_{2}+B\otimes\tau_2]^{-N/2}\ {\rm d}P(A,B){\rm d}r} \nonumber \\
\label{der.1}
\fl&=\lim_{\tilde{\epsilon}\to0}\frac{\int {\rm e}^{\Tr(\epsilon\id_{2N}-\ii A)\widehat{C}}\det[r_A r_A^\intercal+\epsilon\id_{2N}-\ii A+B]^{-N}\ {\rm d}P(A,B){\rm d}r}{\int {\rm e}^{-r_A^\intercal r_A+\Tr(\epsilon\id_{2N}-\ii A)\widehat{C}}\det[\epsilon\id_{2N}-\ii A+B]^{-N}\ {\rm d}P(A,B){\rm d}r}
\end{eqnarray}
with the measure
\begin{eqnarray}
    {\rm d}P(A,B)={\rm e}^{\Tr\Omega B+\tilde{\epsilon}\Tr(\epsilon\id_{2N}-\ii A)^2+\tilde{\epsilon}\Tr B^2}{\rm d}A{\rm d}B.
\end{eqnarray}
We have exploited the two eigenvalues $\pm1$ of $\tau_2$ and the symmetry of $A$ and antisymmetry of $B$ when going from the first to the second line in \eref{der.1}.
The determinant in the numerator can be simplified as follows
\begin{eqnarray}
    \fl\det[r_A r_A^\intercal+\epsilon\id_{2N}-\ii A+B]&=\det[\epsilon\id_{2N}-\ii A+B]\det[\id_{2N}+r_A r_A^\intercal(\epsilon\id_{2N}-\ii A+B)^{-1}] \nonumber \\
    \fl&=\det[\epsilon\id_{2N}-\ii A+B](1+ r_A^\intercal(\epsilon\id_{2N}-\ii A+B)^{-1}r_A)\nonumber \\
    \fl&=\det[\epsilon\id_{2N}-\ii A+B](1+ r_A^\intercal\widetilde{A}r_A/2)\label{det.rewrite}
\end{eqnarray}
with the abbreviation
\begin{equation}
\label{eq:A-tilde-def}
    \widetilde{A}=(\epsilon\id_{2N}-\ii A+B)^{-1}+(\epsilon\id_{2N}-\ii A-B)^{-1}.
\end{equation}
In the last equality we have used $r_A^\intercal Xr_A=r_A^\intercal(X+X^\intercal)r_A/2$ for any matrix $X$. The corresponding integral over $r$, thus, takes the following form
\begin{eqnarray}
\label{eq:r-integral-purity}
\frac{\int_{\mathbb{R}^{2N_A}} (1+ r_A^\intercal\widetilde{A}r_A/2)^{-N}\ {\rm d}r}{\int_{\mathbb{R}^{2N_A}} {\rm e}^{-r_A^\intercal r_A}\ {\rm d}r}&=\frac{2^{N_A}}{\det(\Pi_A\widetilde{A}\Pi_A^\intercal)^{1/2}}\frac{\int_{\mathbb{R}^{2N_A}} (1+ r^\intercal r)^{-N}\ {\rm d}r}{\int_{\mathbb{R}^{2N_A}} {\rm e}^{-r^\intercal r}\ {\rm d}r}\nonumber\\
&=\frac{2^{N_A}}{\det(\Pi_A\widetilde{A}\Pi_A^\intercal)^{1/2}}\frac{\int_{0}^\infty (1+ s)^{-N}s^{N_A-1}\ {\rm d}s}{\int_{0}^\infty {\rm e}^{-s}s^{N_A-1}\ {\rm d}s}\nonumber\\
&=\frac{(N-N_A-1)!}{(N-1)!}\frac{2^{N_A}}{\det(\Pi_A\widetilde{A}\Pi_A^\intercal)^{1/2}},
\end{eqnarray}
where we first substituted $r\to\sqrt{2}(\Pi_A\widetilde{A}\Pi_A^\intercal)^{-1/2}r$ and then went into polar coordinates $r=\sqrt{s}e$ with $s\geq0$ the squared radius and $e$ the unit vector which gives a ratio of a Beta and a Gamma function.
Plugging this back into~\eref{der.1}, we obtain
\begin{eqnarray}
\label{der.2}
\fl\left\langle\Tr\rho_{C_A}^2\right\rangle=&\frac{2^{N_A}(N-N_A-1)!}{(N-1)!}
\\
\fl &\times\lim_{\tilde{\epsilon}\to0}\frac{\int {\rm e}^{\Tr(\epsilon\id_{2N}-\ii A)\widehat{C}}\det[\epsilon\id_{2N}-\ii A+B]^{-N} {\rm d}P(A,B)}{\int {\rm e}^{\Tr(\epsilon\id_{2N}-\ii A)\widehat{C}}\det[\epsilon\id_{2N}-\ii A+B]^{-N}\det(\Pi_A\widetilde{A}\Pi_A^\intercal)^{1/2}\ {\rm d}P(A,B)}
\nonumber
\end{eqnarray}
for which we have to perform the saddle point analysis. We would like to highlight that up to now no approximation has been performed.

The saddle point analysis works along the same lines as for our previous work~\cite{5authors} because we assume $N_A\ll n$ so that the term $\det(\Pi_A\widetilde{A}\Pi_A^\intercal)^{-1/2}$ does not influence the saddle point in the numerator. We are only here considering the mean field approximation and disregard any fluctuation around it. Hence, we only aim at computing the saddle point of the integrals in~\eref{der.2}.

Let us summarise the main steps of~\cite{5authors}. We rewrite first
\begin{eqnarray}\label{saddle.start}
\det[\epsilon\id_{2N}-\ii A+B]^{-N}=\exp\left[-N\Tr\log(\epsilon\id_{2N}-\ii A+B)\right]
\end{eqnarray}
and then differentiate in $B$ and $A$ leading to the saddle point equations
\begin{eqnarray}
\Omega &=\frac{N}{2}\left[(\epsilon\id_{2N}-\ii A+B)^{-1}-(\epsilon\id_{2N}-\ii A-B)^{-1}\right],\label{sad.1}\\
C_{(j)}&=\frac{N}{2}\left[\left\{(\epsilon\id_{2N}-\ii A+B)^{-1}\right\}_{(j)}+\left\{(\epsilon\id_{2N}-\ii A-B)^{-1}\right\}_{(j)}\right],\label{sad.2}
\end{eqnarray}
respectively, with $j=1,\ldots,n$. We recall that the symmetries of $A$ and $B$ need to be respected when differentiating and that $\tilde{\epsilon}$ terms can be neglected in the limit $\tilde{\epsilon}\to0$. In the next step, we define the matrices
\begin{eqnarray}
\label{eq:B-tilde-def}
\fl\widetilde{B}=(\epsilon\id_{2N}-\ii A)^{-1/2}B(\epsilon\id_{2N}-\ii A)^{-1/2} \quad{\rm and}\quad D=(\epsilon\id_{2N}-\ii A)^{1/2}\Omega (\epsilon\id_{2N}-\ii A)^{1/2}\,,
\end{eqnarray}
so that equation~\eref{sad.1} can be recast into
\begin{eqnarray}
\label{eq:D-def}
D=\frac{N}{2}\left[(\id_{2N}+\widetilde{B})^{-1}-(\id_{2N}-\widetilde{B})^{-1}\right]=-N\,\widetilde{B}(\id_{2N}-\widetilde{B}^2)^{-1},
\end{eqnarray}
which is equivalent to $\id_{2N}-\widetilde{B}^2=-N\,\widetilde{B}D^{-1}$.
The latter equation implies that $\widetilde{B}$ and $D$ must commute so that for $\epsilon\approx0$ $B$ must share the block structure of $A$. The solution of the saddle point equations is then given by
\begin{eqnarray}\label{sad.sol}
\fl B_0=\frac{N}{2}\left[(\hat\tau_2-\widehat{C})^{-1}+(\hat\tau_2+\widehat{C})^{-1}\right] \quad{\rm and}\quad A_0=\frac{N}{2\ii}\left[(\hat\tau_2-\widehat{C})^{-1}-(\hat\tau_2+\widehat{C})^{-1}\right].
\end{eqnarray}
Those results can be readily obtained by adding and subtracting Eqs.~\eref{sad.1} and~\eref{sad.2}.

We emphasise that the solutions~\eref{sad.sol} can only be reached when $\widehat{C}$ are positive definite and their symplectic eigenvalues are larger than $1$ as then the matrices $-\ii A\pm B=n(\widehat{C}\pm\tau_2)^{-1}$ are positive definite Hermitian. As in the previous work~\cite{5authors}, the determinants prevent a deformation of the contour to the saddle point in the limit $n\to\infty$ when these conditions are not satisfied. We implement this condition with the help of the matrix Heaviside step function $\Theta(\widehat{C}+\Omega )$ which is unity if the matrix inside is positive definite and vanishes otherwise.

The mean field approximation is then given by
\begin{equation}
\left\langle\Tr\rho_{C_A}^2\right\rangle\approx\frac{2^{N_A}(N-N_A-1)!}{(N-1)!}
\frac{1}{\det(\Pi_A\widetilde{A}_0\Pi_A^\intercal)^{1/2}}
\end{equation}
with
\begin{eqnarray}
   \fl\widetilde{A}_0 &=&(\epsilon\id_{2N}-\ii A_0+B_0)^{-1}+(\epsilon\id_{2N}-\ii A_0-B_0)^{-1}\nonumber\\
   \fl&=& \left(\epsilon\id_{2N}+n(\hat\tau_2+\widehat{C})^{-1}\right)^{-1}+\left(\epsilon\id_{2N}-N(\hat\tau_2-\widehat{C})^{-1}\right)^{-1}=\frac{2}{n}\widehat{C}+\mathcal{O}(\epsilon).
\end{eqnarray}
Thence, the average purity of 
the density matrix in a small subspace over the ensemble of constrained pure Gaussian states is
\begin{equation}
\label{eq:app-final-purity}
\fl\left\langle\Tr\rho_{C_A}^2\right\rangle\approx\frac{2^{N_A}(N-N_A-1)!}{(N-1)!}\frac{1}{\det(\frac{2}{n}\Pi_A\widehat{C}\Pi_A^\intercal)^{1/2}}\approx
\frac{1}{\det(\Pi_A\widehat{C}\Pi_A^\intercal)^{1/2}}=
\Tr\rho_{\widehat{C}}^2,
\end{equation}
where we have exploited $n\gg N_A$. In conclusion, we see that ensemble averaged purity is the same as the purity of a mixed state with the constraint correlation matrix in the same small subspace.

\section{Higher purity of a density matrix at small subsystem size}
\label{sec:average-higher-order}

The general set-up is similar 
to the previous one 
but now considering
$\left\langle\Tr\rho_{C_A}^x\right\rangle$
for $x=3,4,\ldots$
To achieve this we use a coherent state representation
of Gaussian density matrices which in essence can be found, \textit{e.g.}, in
\cite{Serafini}.

\subsection{Displacement operator}
Given $r=(q_1,p_1,\dots,q_N,p_N)$, we define the general displacement operator\footnote{Alternatively, it is also common to write displacement operators as product of individual mode displacements $\mathcal{D}_{\alpha}=e^{\alpha \hat{a}^{\dagger}-\alpha^* \hat{a}}$, where $\alpha=\frac{1}{\sqrt{2}}(q+\ii p)$.}
\begin{equation}
    \mathcal{D}_r=e^{-\ii r^\intercal\Omega\hat{\xi}}\,.
\end{equation}
The displacement operators satisfies the product relation
\begin{equation}\label{displacement.identity}
\mathcal{D}_{r}\mathcal{D}_{r'} = 
e^{\frac{i}{2}r^\intercal \Omega r'}
\mathcal{D}_{r+r'}
\end{equation}
and its trace is a delta distribution in $r$ given by
\begin{equation}
    \mathrm{Tr}[\mathcal{D}_{r}]=(2\pi)^{N}\delta(r)\,.\label{eq:trace-D}
\end{equation}

The characteristic function $\chi(r)$ of any state $\rho$ is defined as
\begin{equation}\label{char.func}
\chi\left({r}\right) = \mathrm{Tr}\left[\mathcal{D}_{r}\,\rho\right]
\end{equation}
and its inverse relation is known as the Fourier-Weyl relation
\begin{equation}
\label{eq:AS-2.16}
\rho =
\frac{1}{(2\pi)^{N}}
\int_{\mathbb{R}^{2N}} {\rm d}r
\,\chi({r})\,
\mathcal{D}_{-r}
\end{equation}
with ${\rm d}r=\prod_{j=1}^{N}{\rm d}q_j{\rm d}p_j$. The normalization can be checked by plugging~\eref{eq:AS-2.16} into~\eref{char.func}, then using~\eref{displacement.identity} and~\eref{eq:trace-D}.

We may restrict to a subsystem consisting of only $N_A$ modes, where we have $r_A=(q_1,p_1,\dots,q_{N_A},p_{N_A})$, $\hat{\xi}_A=(\hat{q}_1,\hat{p}_1,\dots,\hat{q}_{N_A},\hat{p}_{N_A})$ and the appropriate restriction of the symplectic form $\Omega_A$, such that
\begin{equation}
    \mathcal{D}_{r_A}=e^{-\ii r_A^ \intercal\Omega_A\hat{\xi}_A}\,.
\end{equation}

\subsection{Coherent state representation of
$\Tr\rho_{C_A}^x$}

In order to evaluate $\Tr\rho_{C_A}^x$ for a subsystem $A$, we use the replica trick for density operators written in terms of the characteristic function. For a general $\rho$, we express $x-1$ factors in terms of replicas with the help of~\eref{eq:AS-2.16}
\begin{equation}
\Tr
\left[\rho^x\right]
= 
\Tr \left[
\rho
\prod_{j=1}^{x-1}
\frac{1}{(2\pi)^{N}}
\int_{\mathbb{R}^{2N}} {\rm d}\,r^{(j)}
\,\chi\left(r^{(j)}\right)\,
\mathcal{D}_{-r^{(j)}}
\right]
\end{equation}
and then apply~\eref{displacement.identity} $(x-1)$-times,
\begin{eqnarray}
\prod_{j=1}^{x-1}
\mathcal{D}_{-r^{(j)}}
&=& e^{\frac{i}{2}\sum_{a=1}^{x-1}\sum_{b=r+1}^{x-1}(r^{(a)})^\intercal \Omega r^{(b)}} 
\mathcal{D}_{-\sum_a r^{(a)}}.
\end{eqnarray}
The remaining trace can be carried out with the help of~\eref{char.func} which yields the expression
\begin{eqnarray}
\Tr
\left[\rho^x\right]
&= &
\frac{1}{(2\pi)^{N(x-1)}}
\int_{\mathbb{R}^{2N(x-1)}} {\rm d}\,r^{(1)}\cdots{\rm d}\,r^{(x-1)}
\,e^{\frac{i}{2}\sum_{a=1}^{x-1}\sum_{b=r+1}^{x-1}(r^{(a)})^\intercal \Omega r^{(b)}} \nonumber\\
&&\times\chi\left(r^{(1)}\right)\cdots\chi\left(r^{(x-1)}\right)
\chi\left(-\sum_a r^{(a)}\right).
\label{eq:coherent-state-rep-pure-state}
\end{eqnarray}
Up to now this calculation hold for an arbitrary quantum state $\rho$ of $N$ modes.

We now turn our focus to a Gaussian state $\rho_{C_A}$ in a subsystem $A$ of $N_A$ modes whose characteristic function is in terms of ${r}$ equal to
\begin{equation}\label{char.Gauss}
    \chi_{C_A}({r})
    = e^{-\frac{1}{4}
{r}^\intercal\Omega^\intercal  C_A \Omega\, {r}},
\end{equation}
where $C_A=C_A^\intercal>0$ is the covariance matrix defined in~\eref{covariance}.
Combining this with our general considerations leads us to the expression
\begin{eqnarray}
\label{eq:Tr-rho-power-x}
\Tr_A
\left[\rho_{C_A}^x\right]
&= 
\frac{1}{(2\pi)^{N_A(x-1)}}
\int_{\mathbb{R}^{2N_A(x-1)}} {\rm d}\,r^{(1)}\cdots{\rm d}\,r^{(x-1)}
\left[
\prod_{a=1}^{x-1}
\,\chi_{C_A}\left({r}^{(a)}\right)\,\right]\,
\nonumber \\
&
\qquad\times\chi_{C_A}\left(-\sum_b{r}^{(b)}\right)\,
e^{\frac{i}{2}\sum_{1\leq a<b\leq x-1} (r^{(a)})^\intercal \Omega r^{(b)}} .
\end{eqnarray}
It is convenient to 
introduce the following vector and two matrices
\begin{equation}
    e=\left[\begin{array}{c} 
    1\\
    \vdots\\
    1
    \end{array}\right]\in\mathbb{R}^{(x-1)}  \quad{\rm and}\quad
    v=\left[{r}^{(1)},\dots,{r}^{(x-1)}\right]\in\mathbb{R}^{2N_A\times(x-1)}
\end{equation}
and
\begin{equation}
    J=\left(\begin{array}{cccc} 
    0 & 1 &  \cdots & 1 \\
    -1 & 0 &  \cdots & 1 \\
    \vdots & \vdots &   & \vdots \\
    -1 & -1 &  \cdots & 0
    \end{array}\right)\in\mathbb{R}^{(x-1)\times(x-1)}.
\end{equation}
Using~\eref{char.Gauss} and changing the integration variables $v\to\Omega^{-1} v$, the average of the integral in
\eref{eq:Tr-rho-power-x} can then be written as
\begin{eqnarray}
\label{eq:Tr-rho-power-x-2}
\fl\left\langle\Tr_A
\left[\rho_{C_A}^x\right]\right\rangle
&=& 
\left\langle
\frac{1}{(2\pi)^{N_A(x-1)}}
\int_{\mathbb{R}^{2N_A(x-1)}}
dv\,
e^{-\frac{1}{4}\mathrm{Tr}\left[v^\intercal C_Av \right]
-\frac{1}{4}e^\intercal v^\intercal\, C_A\, v e 
+ \frac{i}{4}\mathrm{Tr}\left[Jv^\intercal\Omega v \right]
}\right\rangle.
\end{eqnarray}
For $x=2$, the third term in the exponent drops out as then the matrix $J$ is $1\times 1$ and therefore vanishes. The first and second term in the exponent become equal in this case. Then, the Gaussian integral over $v$ can then be carried out and we arrive at~\eref{purity.start}.

If the measure itself is Gaussian, as it is in our case, we can proceed to switch the order between the averaging and the integral over $v$, and carry out the averaging as Gaussian integrals. This will be our first task in the next subsection.

\subsection{Averaging and stationary phase argument}

Following the arguments of \ref{sec:everage-purity} and starting from~\eref{eq:Tr-rho-power-x-2}, we first need to embed the matrix $v$ into $2N\times (x-1)$ matrix which we do in the very same way as for the single vector $r$ in~\eref{app.A2}, i.e., $v_A=\Pi_A^\intercal v$. Identifying $C_A=\Pi_ASS^\intercal\Pi_A^\intercal$ in
\eref{eq:Tr-rho-power-x-2} and integrating over the symplectic matrix $S$, we need to replace the determinant in the numerator of the second line in~\eref{der.1} as follows
\begin{eqnarray}
  \fl  \det[r_A r_A^\intercal+\epsilon\id_{2N}-\ii A+B]&\longrightarrow&\det\left[\frac{1}{4}v_A\left(\id_{x-1} +e\,e^\intercal\right)v_A^\intercal
+\epsilon\id_{2N}-\ii A+B\right].\label{det.replace}
\end{eqnarray}
Compared to \eref{app.A2}
where $r_A$ is a dummy variable, 
our convention for $r$ here is as in this
appendix, which differs by a factor two; this explains the factor $1/4$ on the right hand side instead of a factor $1/2$. This also means that in the denominator of~\eref{app.A2} we need to replace the vector $r$ by the matrix $v/\sqrt{2}$.

As for purity, the additional term, here 
$\frac{1}{4}{v}_A\left(\id_{x-1} +e\,e^\intercal\right){v}_A^\intercal$,
is a perturbation of rank $x-1$ of the $2N$-dimensional matrix $\epsilon\id_{2N}-\ii A+B$, particularly it is of fixed rank. Hence, we can rewrite the determinant~\eref{det.replace} as in~\eref{det.rewrite} yielding
\begin{eqnarray}
\fl&&\det\left[\frac{1}{4}v_A\left(\id_{x-1} +e\,e^\intercal\right)v_A^\intercal
+\epsilon\id_{2N}-\ii A+B\right]\nonumber\\
\fl&=&\det\left[\epsilon\id_{2N}-\ii A+B\right]\,
\det\left[\id_{2N}
+
\frac{1}{4}
\left(
\epsilon\id_{2N}-\ii A+B\right)^{-1}
v_A
  \left(\id_{x-1} +e\,e^\intercal\right)v_A^\intercal\right]\nonumber\\
\fl&=&\det\left[
\epsilon\id_{2N}-\ii A+B\right]\,
\det\left[\id_{x-1}
+
\frac{1}{4}
v_A^\intercal\left(
\epsilon\id_{2N}-\ii A+B\right)^{-1}
v_A
  \left(\id_{x-1} +e\,e^\intercal\right)\right].
\end{eqnarray}
The first determinant is the same as will appear in the denominator of \eref{der.1}.
There is one more term in the numerator when going over from $r$ to $v$ in~\eref{app.A2}, namely $e^{i\,\mathrm{Tr}\left[Jv^\intercal\Omega v \right]/4}$. It is this term which prevents us to perform the integral over $v$ in a similar way as we have carried out the integral over $r$ in~\eref{eq:r-integral-purity}. Instead we stay with the second determinant  which is taken of a perturbuation of the identity matrix  which is of rank $x-1$. Thus, it will not affect the saddle point analysis which has been carried out after~\eref{saddle.start}.

The mean field approximation to 
$\left\langle\Tr_A
\left[\rho_{C_A}^x\right]\right\rangle$ 
for a fixed value $v$ (fixed value of the $x-1$ replicas of the
coherent state parameters pertaining to subsystem $A$) is hence
found by inserting the variational maximum of the parameter matrices  
$(\epsilon\id_{2N}-\ii A+B)^{-1}\to\frac{1}{n}\left(\widehat{C}+\hat\tau_2\right)$
in above which gives for $n\to\infty$
\begin{eqnarray}
\fl\det\left[\id_{x-1}
+ 
\frac{1}{4n} v_A^\intercal
\left(\widehat{C}+\hat\tau_2\right)
v_A
\left(\id_{x-1} +e\,e^\intercal\right)\right]^{-N}
\rightarrow \exp\left[-\frac{1}{4} \mathrm{Tr}\left[
v_A^\intercal
\widehat{C}
v_A\left(\id_{x-1} +e\,e^\intercal\right) \right]\right].\nonumber\\
\fl
\end{eqnarray}
The part involving $\tau_2$ drops out as this matrix is antisymmetric, \textit{i.e.}, $\mathrm{Tr}\left[v_A^\intercal\hat\tau_2 v_A(\id_{x-1} +e\,e^\intercal)\right]=0$.
%The resulting expression is actually the coherent state representation of the unaveraged trace for the density matrix determined by the constraint  correlation matrix $\widehat{C}_A=\Pi_A\widehat{C}\Pi_A^\intercal\in\mathbb{R}^{2N_A\times 2N_A}$.

Summarising, we have arrived at the approximation
\begin{equation}
\label{eq:app-final-tr-rho-x}
\left\langle\Tr\rho_{C_A}^x\right\rangle\approx
\Tr\rho_{\widehat{C}_A}^x.
\end{equation}
This relation follows from comparing the initial
representation of the higher order purity of a mixed Gaussian state in 
\eref{eq:coherent-state-rep-pure-state}
with the constraint correlation matrix $\widehat{C}_A=\Pi_A\widehat{C}\Pi_A^\intercal$, with the average over pure Gaussian states
\eref{eq:Tr-rho-power-x-2}, as shown above. For completeness, we can evaluate the Gaussian integral in replica space directly, as we will now proceed to do.

\subsection{The coherent state representation as a determinant in replica space}

To resolve the coherent state representation more
explicitly we continue from
\begin{eqnarray}
\fl\left\langle\Tr_A
\left[\rho_{C_A}^x\right]\right\rangle
\approx
\frac{1}{(2\pi)^{N_A(x-1)}}
\int_{\mathbb{R}^{2N_A\times(x-1)}}
{\rm d}{v}\,
e^{-\frac{1}{4} \mathrm{Tr}\left[{v}^\intercal\widehat{C}_A
{v}
\left(\id_{x-1} +e\,e^\intercal\right) \right]
+\frac{\ii}{4}\mathrm{Tr}\left[J{v}^\intercal\Omega {v} \right],
}
\end{eqnarray}
and carry out the Gaussian integral leading to
\begin{equation}
\label{eq:determinant-Photo-2}
\fl\left\langle\Tr_A
\left[\rho_{C_A}^x\right]\right\rangle
\approx  
2^{N_A(x-1)}
\det\left[\widehat{C}_A
\otimes \left(\id_{x-1} +e\,e^\intercal\right)
- \ii\Omega \otimes J 
\right]^{-\frac{1}{2}}.
\end{equation}
For the special case $x=2$ the matrix $J$ is zero, and $\id_{x-1}
=e\,e^\intercal=1$ so that we end up with the result~
\eref{eq:purity-rhoA} in the main body of the paper. Hence, if $\widehat{C}_A=S_A\,\mathrm{diag}(\mu,\mu)\,S_A^\intercal$ (in particular $N_A=1$)
the result is $\mu^{-1}$.

In the general case, we diagonalise $\widehat{C}=\widehat{S}\left(\oplus_j \mu_j\id_2\right)\widehat{S}^\intercal$ with a symplectic matrix $\widehat{S}\in\mathbb{R}^{2N\times 2N}$. If we assume that the restricted covariance matrix $\widehat{C}_A$ respects the block structure of $\widehat{C}$, the restricted matrix $\widehat{S}_A=\Pi_A\widehat{S}\Pi_A^\intercal$ is symplectic, as well, and diagonalises $\widehat{C}_A$. This is actually the case we want to study.

We plug this diagonalisation into~\eref{eq:determinant-Photo-2} where the symplectic matrix $\widehat{S}$ drops out. Thence, the determinant factorises into $N_A$  determinants of size $2(x-1)\times 2(x-1)$, i.e.,
\begin{eqnarray}
    \fl\left\langle\Tr_A
\left[\rho_{C_A}^x\right]\right\rangle
&\approx&2^{N_A(x-1)}\prod_{j\in A}\det \left(\begin{array}{ll}
    \mu_j\left(\id_{x-1} + ee^\intercal\right) & \qquad -\ii J \\
    \qquad \ii J &  \mu_j\left(\id_{x-1} + ee^\intercal\right)
    \end{array}\right)^{-1/2}\nonumber\\
    \fl&=&\frac{2^{N_A(x-1)}}{x^{N_A/2}}\prod_{j\in A}\det \left(
    \mu_j^2\left(\id_{x-1} + ee^\intercal\right)- J\left(\id_{x-1} + ee^\intercal\right)^{-1}J \right)^{-1/2}.
\end{eqnarray}
In the second line, we have applied the Schur complement and exploited the determinant  $\det(\id_{x-1} + ee^\intercal)=x$.

The
inverse of the matrix $\id_{x-1} + ee^\intercal$
is $\id_{x-1} - \frac{1}{x}ee^\intercal$.
%\footnote{
%We also have
%$\left(\id_{x-1} -\frac{1}{x} ee^\intercal\right)^{\frac{1}{2}}=\left(\id_{x-1} - \frac{1}{\sqrt{x}(1+\sqrt{x})}ee^\intercal\right)$.}
Therefore, what remains to be evaluated are the determinants
\begin{eqnarray}
  \Delta_j&=& \det \left(
    \mu_j^2\left(\id_{x-1} + ee^\intercal\right)- J\left(\id_{x-1} -\frac{1}{x} ee^\intercal\right)J \right)%\nonumber\\
    %&=&\det \left(\mu_j^2\id_{x-1}-J^2 + [\mu_je,x^{-1/2}Je]\left[\begin{array}{c}\mu_je^\intercal\\ x^{-1/2}e^\intercal J\end{array}\right] \right).
\end{eqnarray}
Next, we express matrix $J$ in terms of the antiperiodic shift operator
\begin{equation}
J = \sum_{l=1}^{x-2} T^l \qquad{\rm with}\quad
    T = 
    \left[\begin{array}{ccccc} 
    0 & 1 &  0 & \cdots & 0 \\
    0 & 0 & 1 &\cdots & 0 \\
    \vdots & \vdots & \vdots & \vdots  & \vdots \\
    0 & 0 &  \cdots & 0 & 1 \\
    -1 & 0 &  \cdots & 0 & 0
    \end{array}\right]
\end{equation}
Since $T^{x-1}=-\id_{x-1}$ we have
\begin{equation}
\label{eq:J-geometric}
    J=\frac{\id_{x-1}+T}{\id_{x-1}-T}
\end{equation}
which follows from the geometric sum, and the eigenvalues of $T$ are $e^{\ii\pi(2l-1)/(x-1)}$ with $l=1,\ldots,x-1$. This allows us to simplify the expression to
\begin{eqnarray}
  \Delta_j&=&\frac{1}{4}\det \left(
    \mu_j^2(\id_{x-1}-T)^2-(\id_{x-1}+T)^2+\Xi \right)
\end{eqnarray}
with
\begin{eqnarray}
\label{eq:Xi-def}    \Xi&=&\mu_j^2(\id_{x-1}-T)ee^\intercal(\id_{x-1}-T)+x^{-1}(\id_{x-1}+T) ee^\intercal(\id_{x-1}+T)\nonumber\\
    &=&[\mu_j(\id_{x-1}-T)e,x^{-1/2}(\id_{x-1}+T) e]\left[\begin{array}{c} \mu_je^\intercal(\id_{x-1}-T)\\ x^{-1/2}e^\intercal(\id_{x-1}+T)\end{array}\right],
\end{eqnarray}
where we have pulled out the matrix $(\id_{x-1}-T)^{-1}$ from the left and right and used $\det(\id_{x-1}-T)=\prod_{l=1}^{x-1}(1-e^{\ii\pi(2l-1)/(x-1)})=2$
\footnote{This equality can be established by 
noting that the characteristic equation of $T$ ($Tu=\lambda u$) is $u_2=\lambda u_1$,
$u_3=\lambda u_2$\ldots
$u_1=-\lambda u_{x-1}$
which means $\lambda^{x-1}+1=0$.
Writing $-1=e^{i(1+2l)\pi}$
the eigenvalues are 
$\lambda_l=e^{i\frac{1+2l}{x-1}\pi}$, and the product is the characteristic polynomial evaluated at $\lambda=1$.}
In the next step, we pull out the matrix $\Sigma=\mu_j^2(\id_{x-1}-T)^2-(\id_{x-1}+T)^2$. The determinant of this matrix is
\begin{eqnarray}
    \fl\det[\mu_j^2(\id_{x-1}-T)^2-(\id_{x-1}+T)^2]&=&\prod_{l=1}^{x-2}([\mu_j+1]-[\mu_j-1]e^{\ii\pi(2l-1)/(x-1)})\nonumber\\
    &&\times([\mu_j-1]-[\mu_j+1]e^{\ii\pi(2l-1)/(x-1)})\nonumber\\
    \fl&=&([\mu_j+1]^{x-1}+[\mu_j-1]^{x-1})^2.
\end{eqnarray}
Using also the cyclicity of the determinant and \eref{eq:Xi-def} we arrive at
\begin{eqnarray}
\label{eq:Delta-j-two-lines}
\fl&&\Delta_j=\frac{([\mu_j+1]^{x-1}+[\mu_j-1]^{x-1})^2}{4}\\
  \fl&\times&\det \left(\id_{2} + \left[\begin{array}{cc} \mu_j^2e^\intercal(\id_{x-1}-T)\Sigma^{-1}(\id_{x-1}-T)e& \frac{\mu_j}{\sqrt{x}}e^\intercal(\id_{x-1}-T)\Sigma^{-1}(\id_{x-1}+T) e\\ \frac{\mu_j}{\sqrt{x}}e^\intercal(\id_{x-1}+T)\Sigma^{-1}(\id_{x-1}-T)e & x^{-1}e^\intercal(\id_{x-1}+T)\Sigma^{-1} (\id_{x-1}+T) e\end{array}\right] \right).\nonumber
\end{eqnarray}
We note that, using 
\eref{eq:J-geometric},
\begin{eqnarray}
    \fl(\id_{x-1}-T)\Sigma^{-1}(\id_{x-1}+T)&=&(\mu_j^2\id_{x-1}-J^2)^{-1}J=-[(\mu_j^2\id_{x-1}-J^2)^{-1}J]^\intercal
\end{eqnarray}
because $J=-J^\intercal$ is antisymmetric. This implies that the off-diagonals in the $2\times2$ matrix vanish. Furthermore, the diagonals in the $2\times2$ matrix are related by
\begin{eqnarray}
    \fl(\id_{x-1}+T)\Sigma^{-1}(\id_{x-1}+T)&=&J(\mu_j^2\id_{x-1}-J^2)^{-1}J\\
    \fl&&\hspace*{-3cm}=-\id_{x-1}+\mu_j^2(\mu_j^2\id_{x-1}-J^2)^{-1}=-\id_{x-1}+{\mu_j^2}(\id_{x-1}-T)\Sigma^{-1} (\id_{x-1}-T)\nonumber
\end{eqnarray}
which means 
\begin{eqnarray}
e^\intercal(\id_{x-1}+T)\Sigma^{-1}(\id_{x-1}+T)\,e &=&-(x-1) + {\mu_j^2 X}\nonumber
\end{eqnarray}
where
\begin{equation}
    {X = e^\intercal(\id_{x-1}-T)\Sigma^{-1}(\id_{x-1}-T)e}
\end{equation}
is the upper left corner element of the matrix in the second line of  
 \eref{eq:Delta-j-two-lines},
 up to the factor $\mu_j^2$.
The determinant is hence
$\left(1+\mu_j^2 X\right)
\left(1 -\frac{x-1}{x}+\frac{\mu_j^2}{x}X\right)$. 
Thus, we obtain
\begin{eqnarray}\label{Deltaj.exp}
\Delta_j&=&\frac{([\mu_j+1]^{x-1}+[\mu_j-1]^{x-1})^2}{4x}
\,{\left(1+\mu_j^2 X\right)^2}.
%[1+ \mu_j^2e^\intercal(\id_{x-1}-T)\Sigma^{-1}(\id_{x-1}-T)e]^2.
\end{eqnarray}
The vector $(\id_{x-1}-T)e$ is equal to $2e_{x-1}$ where $e_{x-1}$ is $1$ in the $x-1$ entry and zero everywhere else. Similarly, $e^\intercal (\id_{x-1}-T)=2e_1^\intercal$ with $e_1$ the vector with $1$ in the first entry and zero elsewhere. Therefore, we have with the help of the cofactor matrix of $\Sigma$
\begin{eqnarray}
\label{eq:X-from-cofactor}
%   \fl&& e^\intercal(\id_{x-1}-T)\Sigma^{-1}(\id_{x-1}-T)e=
   {X} &=&  
   (-1)^x4\frac{\det\Sigma^{(x-1,1)}}{\det\Sigma}=(-1)^x4\frac{\det\Sigma^{(x-1,1)}}{([\mu_j+1]^{x-1}+[\mu_j-1]^{x-1})^2}
   %,\nonumber\\
   %\fl&&
\end{eqnarray}
with the tridiagonal $(x-2)\times(x-2)$ Toeplitz determinant
\begin{eqnarray}
   \fl \det\Sigma^{(x-1,1)}&=&\det\left[\begin{array}{ccccc}  -2(\mu_j^2+1) & \mu_j^2-1& \ldots&0& 0\\  \mu_j^2-1&-2(\mu_j^2+1) &\ldots&0&0  \\ \vdots& \vdots & &\vdots&\vdots \\ 0&0&\ldots&-2(\mu_j^2+1)&\mu_j^2-1\\ 0&0&\ldots&\mu_j^2-1&-2(\mu_j^2+1) \end{array}\right].
\end{eqnarray}
{We note that the
numerator in the
prefactor in \eref{Deltaj.exp} is the same as the denominator in \eref{eq:X-from-cofactor}, and that the factors of $4$ also match.}
To evaluate the remaining determinant, we define the function
\begin{eqnarray}
    F(t)=1-2\frac{\mu_j^2+1}{\mu_j^2-1}t+t^2=\left(t-\frac{\mu_j+1}{\mu_j-1}\right)\left(t-\frac{\mu_j-1}{\mu_j+1}\right)
\end{eqnarray}
and exploit Ref.~\cite[Corollary III.2]{Kieburg} to find
\begin{eqnarray}
   \det\Sigma^{(x-1,1)}&=&(-1)^x\frac{1}{(x-2)!}(\mu_j^2-1)^{x-2}\left.\partial_t^{x-2}\frac{1}{F(t)}\right|_{t=0}.
\end{eqnarray}
The derivative can be carried out when using
\begin{eqnarray}
   \frac{1}{F(t)}=\frac{\mu_j^2-1}{4\mu_j} \left[\frac{1}{(\mu_j-1)/(\mu_j+1)-t}-\frac{1}{(\mu_j+1)/(\mu_j-1)-t}\right]
\end{eqnarray}
which leads to
\begin{eqnarray}
   \fl \det\Sigma^{(x-1,1)}&=&(-1)^x\frac{(\mu_j^2-1)^{x-1}}{4\mu_j}\left[\frac{(\mu_j+1)^{x-1}}{(\mu_j-1)^{x-1}}-\frac{(\mu_j-1)^{x-1}}{(\mu_j+1)^{x-1}}\right]\\
   \fl&=&(-1)^x\frac{(\mu_j+1)^{2x-2}-(\mu_j-1)^{2x-2}}{4\mu_j}.
\end{eqnarray}
Inserting this back into~\eref{Deltaj.exp} we arrive at
\begin{eqnarray}\label{Deltaj.exp-2}
\fl\Delta_j&=&\frac{[([\mu_j+1]^{x-1}+[\mu_j-1]^{x-1})^2+(-1)^x 4\mu_j^2\det\Sigma^{(x-1,x-1)}]^2}{4x([\mu_j+1]^{x-1}+[\mu_j-1]^{x-1})^2}\\
\fl&=&\frac{[([\mu_j+1]^{x-1}+[\mu_j-1]^{x-1})^2+ \mu_j[(\mu_j+1)^{2x-2}-(\mu_j-1)^{2x-2}]]^2}{4x([\mu_j+1]^{x-1}+[\mu_j-1]^{x-1})^2}\nonumber\\
\fl&=&\frac{[2[\mu_j^2-1]^{x-1}+ (\mu_j+1)^{2x-1}-(\mu_j-1)^{2x-1}]^2}{4x([\mu_j+1]^{x-1}+[\mu_j-1]^{x-1})^2}.\nonumber
\end{eqnarray}
This means for the averaged higher purity
\begin{eqnarray}
    \fl\left\langle\Tr_A
\left[\rho_{C_A}^x\right]\right\rangle
&\approx&2^{N_Ax}\prod_{j\in A}\frac{[\mu_j+1]^{x-1}+[\mu_j-1]^{x-1}}{2[\mu_j^2-1]^{x-1}+ (\mu_j+1)^{2x-1}-(\mu_j-1)^{2x-1}}.
\end{eqnarray}
Finally noting 
\begin{eqnarray}
\fl\left((\mu_j+1)^x - (\mu_j-1)^x\right)
\left[(\mu_j+1)^{x-1}+(\mu_j-1)^{x-1}\right] && \nonumber\\
\fl= (\mu_j+1)^{2x-1}
-(\mu_j-1)(\mu_j^2-1)^{x-1}
+(\mu_j+1)(\mu_j^2-1)^{x-1}
- (\mu_j-1)^{2x-1} && \nonumber \\
\fl= (\mu_j+1)^{2x-1}
+2(\mu_j^2-1)^{x-1}
- (\mu_j-1)^{2x-1} &&
\end{eqnarray}
we find
\begin{eqnarray}
    \left\langle\Tr_A
\left[\rho_{C_A}^x\right]\right\rangle
&\approx& \prod_{j\in A}\frac{2^x}{(\mu_j+1)^x - (\mu_j-1)^x}
\end{eqnarray}
which is
\eref{eq:trace-rhoCii-power-x}
in the main body of the paper, for a factorized mixed Gaussian state fully characterized by its marginals.

\section*{References}
\bibliography{PageProblem.bib,GaussianHawking.bib}
\bibliographystyle{iopart-num}
\clearpage

\end{document}